\documentclass[prd,aps,superscriptaddress,nofootinbib,amsmath,amssymb,twocolumn]{revtex4-1}
\usepackage[dvips]{graphicx}
\usepackage{amsmath,amssymb,latexsym}

\usepackage{amssymb,amsmath,epsfig,bm}
\usepackage[usenames,dvipsnames]{color}
\usepackage{mathrsfs}
\usepackage[bookmarks]{hyperref}

\usepackage{amssymb,amsmath,epsfig,bm}

\newcommand{\rb}{\bar{r}}
\newcommand{\mb}{m\ell}
\newcommand{\Rb}{\bar{R}}

\def\be{\begin{equation}}
\def\ee{\end{equation}  }
\def\bea{\begin{eqnarray}}
\def\eea{\end{eqnarray}  }

\begin{document}          

\title{
Dynamics and Observational Signatures of Shell-like Black Hole Mimickers
}
\date{\today}

\author{Ulf Danielsson}
\affiliation{Uppsala University, Uppsala, Sweden}

\author{Luis Lehner}
\affiliation{Perimeter Institute for Theoretical Physics, 31 Caroline St., Waterloo, ON, N2L 2Y5, Canada}

\author{Frans Pretorius}
\affiliation{Princeton University, NJ, USA}

\begin{abstract}
We undertake the task of studying the non-linear dynamics of quantum gravity motivated
alternatives to black holes that in the classical limit appear as ultra-compact shells
of matter. We develop a formalism that should be amenable to numerical solution
in generic situations. For a concrete model we focus on the spherically symmetric AdS black 
bubble --- a shell of matter at the Buchdahl radius separating a Schwarzschild exterior
from an AdS interior. We construct a numerical code to study the radial dynamics
of and accretion onto AdS black bubbles, with exterior matter provided by scalar fields.
In doing so we develop numerical methods that could be extended to future studies
beyond spherical symmetry. Regarding AdS black bubbles in particular, we find that
the original prescription for the internal matter fluxes needed to stabilize the
black bubble is inadequate in dynamical settings, and we propose a two parameter generalization of
the flux model to fix this. To allow for more efficient surveys
of parameter space, we develop a simpler numerical model adapted to spherically symmetric bubble dynamics.
We identify regions of parameter space that do allow for
stable black bubbles, and moreover allow control to a desired end-state after an accretion
episode. Based on these results, and evolution of scalar fields on black bubble backgrounds,
we {\em speculate} on some observational consequences if what are currently
presumed to be black holes in the universe were actually black bubbles.
\end{abstract}

\maketitle

\tableofcontents

\section{Introduction}
Since the recognition of black hole entropy and formulation of the black hole information paradox,
many efforts have focused on interpreting and reconciling such puzzling
aspects of black holes. These, and related attempts to tame the singularities inside black holes,  
motivated consideration of ``extensions'' to black holes, altering their
structure in the vicinity of the classical horizon and its interior.
This has resulted in proposals for objects like fuzz balls~\cite{Lunin:2001jy}, 
gravastars~\cite{Mazur:2001fv}, black bubbles~\cite{Danielsson:2017riq}, etc.
--- see~\cite{Cardoso:2019rvt} for a review, which also describes many other
exotic compact object (ECO) alternatives to black holes not necessarily
motivated by quantum gravity considerations.

Recently, the ability to detect gravitational waves from merging stellar-mass compact object binaries
(e.g.~\cite{Abbott_2019,theligoscientificcollaboration2021gwtc21}), as well as to observe
horizon scale physics of supermassive black holes (\cite{2019ApJ...875L...1E}), has tremendously
energized this field, motivating the search for observational consequences of ECO
alternatives to Kerr black holes.
However, the majority of existing models of ECOs have not been studied in dynamical, nonlinear settings,
and for some it is even uncertain how that might be done in theory. 
This is clearly an unfortunate state of affairs, even beyond the obvious need for predicting waveforms
in mergers, as understanding the dynamical stability of isolated objects, or
lack thereof, could eliminate some models, and guide refinements to more viable ones.

Here, we focus on a sub-class of ECO modelled as an ultra-compact thin-shell; a 2-sphere surface layer of matter
close to the would-be horizon of the analogous black hole, but still at a macroscopic distance outside,
 that separates a non-singular interior spacetime from the exterior, asymptotically flat spacetime.
To realize the ultimate goal of studying the merger of two such objects, one must resort to numerical simulations. 
This requires the introduction of novel ideas and methods to deal with the new
ingredients such an ECO would bring to a traditional numerical general relativity code :
singular surface layers, matter
fields (including a cosmological constant) confined to the surface and or exterior/interior
spacetimes, new interactions between traditional matter fields and the surface, etc.

Before such novel techniques can be investigated, it is essential to begin
with a viable ECO model. This requires both a classically well-posed problem of the
spacetime and matter system at hand\footnote{Requiring mathematically sound equations together with
suitable initial and boundary conditions.}, and that the ECO solutions are dynamically stable. A promising
candidate in this regard, that we will adopt\footnote{This is a convenient choice, though lessons derived in
our studies are applicable to other models as well.}, are the Anti de-Sitter (AdS) black bubbles proposed by Danielsson, Dibitetto
and Giri~\cite{Danielsson:2017riq}.
This model has so far only been developed for non-rotating or 
slowly rotating spacetimes\cite{Danielsson:2017pvl}. In the non-rotating case,
a thin shell of matter in equilibrium at the Buchdahl radius separates an interior
AdS spacetime from the exterior Schwarzschild spacetime.
The matter, inspired by string theory constructions, consists of a relativistic gas attached to
a membrane, with internal interactions between the components designed to react 
to external perturbations so as to keep the bubble stable. Another goal of this
work then is to explore the stability of these bubbles beyond the quasi-stationary,
linear regime investigated in~\cite{Danielsson:2017riq}.

As a first step toward an ultimate goal of exploring black bubble mergers, we will restrict attention 
to single, spherically symmetric black bubbles. In spherical symmetry one can
adapt the problem to the symmetry, avoiding
many complications one would need to address in a generic scenario.
However, we have intentionally tried to {\em not} do that as much as possible, often complicating the problem
simply for the sake of introducing a feature that would 
be present in a non-symmetric case. This includes {\em not} explicitly imposing the Israel
junction conditions~\cite{1966NCimB..44....1I} (that in spherical symmetry by themselves
can uniquely determine the shell dynamics and map between interior/exterior 
spacetimes),
choosing a metric ansatz where we have gauge waves that propagate at the speed
of light, and using a scalar field as a proxy for gravitational wave
interactions with the shell. 

In a sense, we have been successfull in 
implementing this Einstein-Klein-Gordon-Hydrodynamic (EKGH) model.
However, when first applying it to AdS black bubbles,
we found it failed to address the question of physical stability 
in the large-bubble limit, of interest for astrophysical applications. 
This turns out in part to be due to some of these ``complicating'' choices
we made for the EKGH code, but also in part due to the physics of black bubbles
in the large mass limit. To answer the stability question, which would
be crucial to do before either improving the EKGH code,
or to go beyond spherical symmetry, we here also introduce
a simpler, spherically-symmetric adapted model that can investigate
some aspects of the non-linear, dynamical stability of large black bubbles.

The rest of the introduction outlines the remainder of the paper,
and summarizes the main results. 

In Sec.~\ref{adsblackbubbles} we review aspects
of AdS black bubbles, and give a general formalism to describe such 2+1D matter embedded in 
a dynamical 3+1D spacetime. For surface matter we consider the combination
of fluids proposed in~\cite{Danielsson:2017riq}, though allow for the possibility
of viscosity to be present. For external matter, we consider two scalar fields:
the first does not directly interact with the matter intrinsic to the shell, and can freely propagate across
its surface (i.e., the proxy for gravitational waves), while the second can be absorbed
by the shell to model accretion.

In Sec.~\ref{sec_ss} we specialize to spherical symmetry. First,
in Sec.~\ref{EKG_formalism}, we describe
the full EKHG version of the equations, including our ansatz for the metric,
the resultant evolution equations, constraint equations, initial conditions
and boundary conditions.
In Sec.~\ref{diss_model} we describe the simplified model that can 
explore the dynamics of an AdS black bubble perturbed by an unspecified external
source (i.e. it does not include the gravitational wave proxy scalar field,
and cannot relate the perturbing source to a particular external scalar field profile).

Stability in the AdS black bubble model is achieved via an internal flux
between the gas and brane components. In Sec.~\ref{AdSBubble} we discuss this
in more detail, including the extensions beyond that of the original
model we introduce here. As outlined there, with more details and analysis
given in Appendices \ref{dynamicalproperacc} and \ref{vac_sd}, the flux
prescription of \cite{Danielsson:2017riq} does not result
in stable bubbles if the full dynamical problem is considered, and one demands
the internal flux can only react to local changes in the environment. 
The modifications we have introduced here are somewhat {\em ad hoc},
though our reasoning is if we can identify flux prescriptions that 
lead to stable bubbles it will help guide searches for more fundamental
physical mechanisms that can achieve similar effects.

In Sec.~\ref{implementation} we discuss numerical implementation details
of the spherically symmetric equations given in Sec.~\ref{sec_ss}.
We focus on novel aspects pertaining to this problem, including a weak-form
integration procedure to deal with the singular surface, and a dual 
coordinate scheme to keep the bubble at a fixed location within the computational
grid. More technical details of this are relegated to Appendices \ref{map_rx_A}
and \ref{sec_wfs}.

In Sec.~\ref{results} we give results from evolution of perturbed
black bubbles. In Sec.~\ref{diss_model_results} we focus on the physics
of AdS black bubble dynamics, 
giving examples using flux parameters (guided by the linear analysis 
presented in Appendix~\ref{vac_sd}) that allow for stable, large black bubbles.
In Sec.~\ref{diss_EKG_results} we discuss the limitations of the EKGH
code in this regard; in particular, the two (likely related) problems
are the challenge to achieve sufficient accuracy over multiple light-crossing
times, and a ``mass amplification'' effect
that occurs due to the purely gravitational interaction of scalar
field energy crossing from the exterior to interior
spacetimes. However, the EKGH code is capable of modeling the long term
interactions of the scalar field on a {\em fixed} black bubble background; in this
section then we also present some results for the case of the gravitational wave
proxy field that can freely cross the bubble surface. This suggests some 
remarkable potential observational consequences following black bubble
formation, in particular a slow, nearly monochromatic release of the
energy at the fundamental oscillation frequency of the interior
AdS spacetime, redshifted to near the characteristic frequency of the exterior
black bubble spacetime. However, as estimated in Appendix~\ref{app_loss_rate},
if the AdS lengthscale is set by Planck scale physics, the energy
release will be much too slow to be of relevance for astrophysical sized (stellar 
and supermassive) black holes. Discriminating between black bubble and
black hole mergers would then seem to require understanding the prompt
signal following a collision, or unusual interior physics/anomalously large
lengthscales; we speculate on these topics, as well as give directions
for future work in Sec.~\ref{discussion}.

\section{AdS black bubbles and General considerations}\label{adsblackbubbles}
Among suggested alternatives to black holes resulting from gravitational collapse
are {\em ``black bubbles''}, first proposed in~\cite{Danielsson:2017riq}.
Though their creation is argued to come from a quantum tunneling process,
and their surface structure to be composed of string-theory
inspired matter and higher dimensional geometric constructions,
once formed, astrophysically large black bubbles can to good approximation
be described by classical physics. In this limit they are solutions
to 4 dimensional Einstein gravity coupled to matter--- an electromagnetic
gas, a membrane, and a subleading stiff gas ---confined to a thin shell. For the non-rotating case, the interior
is AdS spacetime with a cosmological constant $\Lambda\equiv-\ell^2$,
the exterior is Schwarzschild with a gravitational (ADM) mass $m$.
For both stability and to possess an equation of state
that ``naturally'' follows from the string theory construction, 
the shell is required to be at the Buchdahl radius $r=9m/4$. 

Of particular interest is to assess whether a black bubble can be
regarded as a viable candidate for an ``exotic''
alternative to a black hole. Notably, this would mean confronting the behavior of merging
black bubbles with gravitational wave observations of what are currently
interpreted as merging black holes~\cite{LIGOScientific:2016aoc,LIGOScientific:2020ibl}, and 
whether accretion onto supermassive
black bubbles is distinguishable from that onto supermassive black holes as observed
by the Event Horizon Telescope~\cite{EventHorizonTelescope:2019dse}.

Here, we focus on working toward the goal of using gravitational
wave observations to study the viability of black bubbles.
This requires that the classical system admits a well-posed initial value problem
outside of the static, spherically symmetric spacetime ansatz where they
were first analyzed, and that single black bubbles are dynamically stable
to generic perturbations. From the classical perspective this would
include non-linear perturbations, at least as long as the energy
of the perturbation is not large enough to require considering it
a ``creation event'', rather than a perturbation\footnote{That also begs the question of
whether classical physics can even approximately address the coalescence phase
of bubble mergers, especially in the comparable mass case where there would
be a significant change in the mass of the final bubble compared to either
progenitor. We leave that to future work to contemplate.}. 
In~\cite{Danielsson:2017riq} a first step toward addressing the stability
question was taken, where it was argued that several ingredients are necessary
for black bubbles to be stable under radial perturbations. The primary ones are
that the gas comprising the bubble must be at the local Unruh (acceleration) temperature as measured
by a comoving observer just outside the shell, and that an internal flux
between the relativistic gas and membrane components of the shell operates
to react to perturbations to instantaneously maintain this temperature. 

In this work, we take a couple of additional steps toward the goal of assessing the
ultimate (classical) viability of black bubbles. The first is to study the
stability of spherically symmetric bubbles undergoing dynamical
radial perturbations, not necessarily small, excited by some external
agent. As we show in Appendices \ref{dynamicalproperacc} and \ref{vac_sd},
the original kinematic stability analysis of~\cite{Danielsson:2017riq} missed a dynamical
component of the 4-acceleration that feeds into the flux, the latter part
of which has a destabilizing effect on the black bubble. Thus the original
flux prescription does not leat to stability, and in Sec.~\ref{AdSBubble} we offer extensions
to it that can lead to radially stable bubbles. 

The second step is to formulate the problem in a manner 
that does not rely on spherical symmetry, even though
our example implementation is restricted to it.
The technical issue here is how to deal with
singular (delta function) distributions of matter coupled
to the Einstein equations in a situation without symmetries.
In particular, in general the shell world tube cannot be considered a spacetime boundary
in a mathematical (or physical) sense where boundary conditions need to be applied; for example, gravitational
waves can freely cross this location, and any influence the matter might exert
on the gravitational waves is governed by the Einstein equations,
not any ``boundary condition'' one places there. 
Of course, this is exactly where the Israel (sometimes also referred to as the Lanczos-Darmois-Israel-Sen)
junction conditions come from, but in spherical symmetry one {\em can} effectively
employ them as boundary conditions for the spacetime on either side of the world tube
(essentially because there are no gravitational waves in spherical symmetry). 
However, it is difficult to envision how such an approach could be
extended to spacetimes without symmetries, in particular where
the surface layer might not be the dominant source of curvature
(for example, it would have to work in the limit of a ``fictitious surface'' 
where the stress-energy of the surface goes to zero). 
Instead, as described next, we
adopt a first principles approach, adding a distributional source for the shell matter to the
Einstein equations, arriving at the junction conditions as a consequence rather than a condition
put in {\em a priori}.

\subsection{Formulation}
For our target model, there are $3$ distinct regions. An interior region, with a non-zero (negative)
cosmological
constant, an exterior region with $\Lambda=0$, and a shell that separates them. The shell,
with a non-trivial stress energy tensor composed of several matter components outlined
below, provides the physical mechanism that can, in principle, stabilize
the shell. In addition, we add bulk scalar fields to model dynamical scenarios, both to
perturb the black bubble via a gravitational interaction, and a direct
interaction where the black bubble accretes scalar field energy.
To account for all these ingredients, on a rather general footing, we proceed as follows.
We consider the Einstein equations in the full spacetime (using units where Newton's
constant $G=1$ and the speed of light $c=1$)
\be\label{efe}
G_{ab}=8\pi T_{ab},
\ee
with a stress-energy tensor of the form 
\be\label{Tab_net}
T_{ab}=   ^{(\rho)}T_{ab}\ \delta(s) 
        + ^{(\psi)}T_{ab} 
        + ^{(\xi)}T_{ab}\ \Theta(s) 
        - \frac{\Lambda}{8\pi} g_{ab}\ \Theta(-s).
\ee
The net stress-energy tensor of the material comprising the shell is $^{(\rho)}T_{ab}$,
$^{(\psi)}T_{ab}$ is that of a bulk scalar field that does not directly
interact with shell material (but can cross the shell location from the outside to inside
and vice-versa), while $^{(\xi)}T_{ab}$ is that of an exterior-only scalar field
that interacts with the shell via appropriately chosen boundary conditions, discussed below.
The shell world tube is described by the level set scalar function $s(x^a)=0$,
with $\delta(s)$ the Dirac delta distribution, and $\Theta(s)$ is the Heaviside step function.
Immediately adjacent to any point on the shell we will normalize $s$ to measure
proper distance orthogonal to the shell at that point, with $s>0$ ($s<0$) on the outside (inside).
The gradient $s_a \equiv s_{,a}$, dual to the vector $s^a = g^{ab} s_b$ normal
to the world tube, will thus be unit ($s^a s_a=1$), and defines the projection tensor
$h_{ab} = g_{ab} - s_a s_b$ onto the shell, as well as the extrinsic
curvature
\be\label{K_def}
K_{ab}=- h^c{}_a h^d{}_b \nabla_c s_d\ 
\ee
evaluated on either side
of the shell as used in the Israel junction conditions. 

\subsection{Scalar field and shell material}
The stress energy tensors for the scalar fields are
\bea\label{sf_set}
^{(\psi)}T_{ab} = \nabla_a \psi \nabla_b \psi - \frac{1}{2}g_{ab}|\nabla \psi|^2,\\
^{(\xi)}T_{ab} = \nabla_a \xi \nabla_b \xi - \frac{1}{2}g_{ab}|\nabla \xi|^2.
\eea
Following~\cite{Danielsson:2017riq} we will build the shell from three fluid components : 
a relativistic gas, a brane and a subleading stiff gas. There, all the fluids were
modeled as perfect (ideal) fluids; here we allow for viscosity
to model dissipative effects. 
The latter is important to account for the entropy growth of the bubble
as it interacts with its environment. Since the gas dominates the entropy
of the bubble, for simplicity then we only add dissipation to that 
component of the shell. To do so we employ the formulation of viscous relativistic hydrodynamics
which only modifies the fluid description to first order in a gradient expansion~\cite{Bemfica_2020,Bemfica_2018,Kovtun_2019}. 
Under certain conditions (including spherical symmetry) we can consider a single 4-velocity $u^a$ to characterize
the flow of all fluid elements, and for simplicity of notation we will do that here.
The resulting stress energy tensor for the shell is
\bea
^{(\rho)}T_{ab}   &\equiv& ^{(\rho_g)}T_{ab} + ^{(\rho_s)}T_{ab} + ^{(\rho_\tau)}T_{ab}\label{rho_set}, \\
^{(\rho_g)}T_{ab} &=& (\rho_g + \mathcal{A}) u_a u_b + (p_g + \Pi) \Delta_{ab},\label{rho_g_set}\\
^{(\rho_s)}T_{ab} &=& \rho_s u_a u_b + p_s \Delta_{ab}\label{rho_s_set},\\
^{(\rho_\tau)}T_{ab} &=& \rho_\tau u_a u_b + p_\tau \Delta_{ab}\label{rho_t_set},
\eea
where $\Delta_{ab}=h_{ab}+u_a u_b$, $\rho_g,\rho_s$ and $\rho_\tau$ are the (equilibrium)
rest-frame energy densities of the gas, string and brane components,
with corresponding pressures $p_g=\rho_g/2,p_s=\rho_s$ and $p_\tau=-\rho_\tau$, respectively.
The viscous modifications to the gas stress energy are captured by $\mathcal{A}$
and $\Pi$, defined as 
\bea
\mathcal{A}&=&\tau_e \left[u^a D_a\rho_g + (\rho_g + p_g) D_a u^a\right]\label{A_vis},\\
\Pi&=&-\zeta D_a u^a + \tau_p \left[u^a D_a\rho_g + (\rho_g + p_g) D_a u^a\right]\label{PI_vis},
\eea
where $\tau_e, \tau_p$ and $\zeta$ are transport coefficients that 
are functions of $\rho_g$, and $D_a\equiv h^b{}_a \nabla_b$.
In general there are additional terms
proportional to the shear tensor $\pi^{ab}$ and heat flux $\mathcal{Q}{^a}$, but these
vanish in spherical symmetry, so we drop them here for simplicity.
$\zeta$ is the bulk viscosity coefficient; ($\tau_e, \tau_p$) are often ignored
in relativistic hydrodynamics, though are required in the first order theory for causality and
to allow for defining (locally at least) well-posed problems\footnote{Though this well-posedness result is
obtained in a somewhat weaker sense than the traditional Sobolev criteria and with
a non-unique entropy current, it has shown promising results in incipient applications~\cite{Pandya:2021ief}.}~\cite{Bemfica_2020,Bemfica_2018,Kovtun_2019}.

\subsection{Matter equations of motion}

For the scalar fields, we impose the usual massless wave equations
\bea
\Box \psi &=& 0,\label{psi_eom_b} \\
\Box \xi &=& 0.\label{xi_eom_b}
\eea
For the shell, the equations of motion stem from net stress energy conservation.
We will further demand that each component of the shell individually
satisfies conservation of its respective stress energy tensor, except 
for an external source term $J^a$ to the gas component to allow for energy exchange
with the scalar field $\xi$, and an internal flux $j^a$ between the gas and brane components
(discussed more in Sec.~\ref{AdSBubble}):
\bea
D_b\ {}^{(\rho_g)}T^{ab} &=& J^a - j^a,\label{rho_g_div}\\
D_b\ {}^{(\rho_s)}T^{ab} &=& 0,\label{rho_s_div}\\
D_b\ {}^{(\rho_\tau)}T^{ab} &=& j^a.\label{rho_t_div}
\eea
Note that we do not explicitly add a source term
to $\xi's$ equation of motion (\ref{xi_eom_b}), as at the location
of the shell we do not impose (\ref{xi_eom_b}), but rather must specify boundary conditions for 
$\xi$ there, and this will effectively compensate for $J^a$.
In spherical symmetry, the only gradients that will
be relevant for these boundary conditions are those in the $u^a$ and $s^a$ directions, and $J^a$,
being intrinsic to the shell, can then only
have a component in the $u^a$ direction. So to simplify the
expressions below, we  define 
\bea
J_U  &\equiv& u_c J^c,\\
\xi_S&\equiv& s^a \nabla_a\xi, \\
\xi_U&\equiv& u^a \nabla_a\xi.
\eea

For consistency with the Einstein equations the net stress energy tensor must satisfy
\be\label{continuity}
T^{ab}{}_{;b}=0.
\ee
Evaluating
this at the shell, substituting in the equations of motion wherever possible,
and averaging $\nabla_a s_b, \nabla_a u_b$ related terms that are discontinuous across the
shell\footnote{The averaging can be justified by integrating the equations in a small
volume about the shell, and taking the limit of the volume to zero; see~\cite{1966NCimB..44....1I}.},
we can split the result into a piece tangent to the shell
\be\label{tab_t}
T^{ab}{}_{;b}\ u_a = \xi_U \xi_S + J_U = 0,
\ee
and one orthogonal to the shell
\be\label{tab_o}
T^{ab}{}_{;b}\ s_a = \frac{\xi_U^2 + \xi_S^2}{2}
+\left[(\rho+P)\tilde{a} + P \tilde{K}\right] + 
\frac{\Lambda}{8\pi} =0.
\ee
In the above, a tilde $(\tilde{\ })$ denotes the averaging, we have defined
\bea
\rho&\equiv&\rho_g+\mathcal{A}+\rho_\tau+\rho_s,\label{rho_net}\\
P&\equiv& p_g+\Pi+p_\tau+p_s,\label{P_net}
\eea 
$K$ is the trace of the extrinsic curvature (\ref{K_def}),
and $a$ is the radial acceleration of $u^a$:
\be\label{four_a_def}
a \equiv \nabla_a u^b u^a s_b.
\ee
The orthogonal piece (\ref{tab_o}) gives the equation of motion for the shell,
while the tangential piece (\ref{tab_t}) constrains the interaction between the gas and
scalar field:
\be\label{tab_t_1}
J_U = - \xi_U \xi_S.
\ee
With this relation in hand, one immediately sees that a pure
Dirichlet ($\xi_U=0$) or Neumann ($\xi_S=0$) boundary condition forces
$J_U=0$. These options effectively implement a reflection with no direct energy exchange 
to the gas (though kinetic energy will still be exchanged via (\ref{tab_o})).
To have the black bubble mimic a black hole and always absorb energy,
one can demand $J_U<0$ (see (\ref{continuity_g}) below).
An obvious choice for this, that we will use in the results presented later, is
\be\label{abs_bc}
\xi_U=\xi_S.
\ee
This is the analog of an ingoing radiation condition at the shell, assuming $\xi$ takes 
the form $\xi\sim \xi(t+r)$ there. Note in this case, the energy associated to the field $\xi$ is
absorbed by the gas.

\section{Restricting to Spherical Symmetry}\label{sec_ss}
So far, we have kept the presentation general, discussing in broad strokes
the governing equations from a global spacetime point of view. To simplify
the form of the equations, in a couple of instances we already imposed 
restrictions consistent with spherical symmetry,
though this did not change the basic structure of the equations
(in particular using a single 4-velocity $u^a$
to describe the shell trajectory and all local fluid velocities,
and only considering energy exchange with the scalar field in this same direction).
Here, we do specialize the equations to spherical symmetry. 

First, in Sec.~\ref{EKG_formalism}, we consider the full Einstein-Klein-Gordon-Hydrodynamic (EKGH)
system of equations in spherical symmetry, giving a set of 1+1D partial differential equations (PDEs) to solve for the spacetime metric
and scalar fields, and a set of ordinary differential equations (ODEs)
for the position and fluid properties of the shell. However, as discussed
in Sec.~\ref{diss_EKG_results}, in the large mass limit of relevance
for astrophysical black holes, the corresponding code for this system
of equations is not adequate to study black bubble evolutions
for the time needed to ascertain their stability. This is in part
due to the large disparity of scales in the problem in that limit,
as well as decisions we made toward the longer term goal of extending
the code beyond spherical symmetry (i.e., if going beyond
spherical symmetry was not of interest, choices better adapted
to the problem could be made, in particular with regards
to the interior and exterior coordinate charts). 

Though just as crucial
as having the correct tools to go beyond spherical symmetry,
is knowing that this is a sensible endeavor in the first place.
Early cases examined with the EKGH code indicated that the original black bubble
prescription is not dynamically stable, leading us to consider the additional
ingredients added to the model described in this paper. However,
a linear stability analysis including the new flux options and dissipation
(Appendix~\ref{vac_sd}) shows that the stability properties can be very
different for small bubbles (relative to the AdS lengthscale $1/\ell$) versus large
bubbles. Thus it would be suspect to use any conclusions of non-linear stability
obtained with the EKGH code in the small bubble case to decide whether it would be worth the considerable
effort needed to first resolve its problems with large bubbles, and
then extend beyond spherical symmetry. Therefore, we implemented a simplified
ODE model, described in Sec.~\ref{diss_model}, that allows us to explore
non-linear stability in the large mass limit, albeit without completely
general scalar-field interactions as allowed by the EKGH system.

\subsection{Einstein-Klein-Gordon-Hydrodynamic system}\label{EKG_formalism}

The formalism described above does not require a coordinate chart that gives
a continuous metric across the shell. Discontinuous charts are convenient
in certain respects, and would be simple
to implement in a code in spherical symmetry. Though again
we want to use methods that could be extended beyond
spherical symmetry in a straightforward manner;
to that end, we will use a metric ansatz which is continuous across the shell. 
Moreover, we will adopt the following ``light-like'' ansatz for the metric so that the equations of motion bear
close resemblance to the typical structure encountered in 3+1D scenarios:
\be
ds^2=e^{2 B(r,t)} (-dt^2 + dr^2) + r^2 e^{2 C(r,t)} d\Omega^2.\label{metric}
\ee
Beyond the obvious structure of the $r-t$ sector of the metric, what 
this light-like ansatz does is allow one to write
the Einstein evolution equations for $B$ and $C$ so that the
principal parts of each are wave equations.

\subsection{Evolution equations}
Let $f(x^a)=r-R(t)$, so the shell is at $f=0$, i.e. at $r=R(t)$. Then $s_a=\nabla_a f/|\nabla f|$,
the coordinate velocity of the shell is $V\equiv dR/dt$, 
$u^a=\gamma(1,V,0,0)$ and $s^a=\gamma(V,1,0,0)$, with $\gamma=e^{-B}/\sqrt{1-V^2}$.
To deal with the
distributional matter of the shell in the numerical code,
we will integrate the equations at the location of the shell
in ``weak-form'', as described in more detail in Sec.~\ref{sec_weak}. This involves integrating
over a volume in the coordinate $r$. In the covariant form of the stress
tensor (\ref{Tab_net}), the shell is a distribution in $s$, and to adapt 
to the integration in $r$ we use
$\int \delta(s) dr$ = $\int \delta(s) (dr/ds) ds$ = $dr/ds|_{s=0}$.
So in anticipation of that, in the equations below $\delta(s)$ has been replaced
with $\delta(f) dr/ds$, where $dr/ds=e^{-B} \sqrt{1-V^2}$, and defining 
$\int \delta(f) dr=1$.
We will write the Einstein and scalar field equations in first order form,
defining $z_t\equiv z_{,t}$ and $z_r\equiv z_{,r}$ for a variable $z(t,r)$.
Then, evolution equations for $B,C$ from the Einstein
equations (\ref{efe}), the wave equations for $\psi$ (\ref{psi_eom}) and $\xi$ (\ref{xi_eom}), 
conservation equations (\ref{rho_g_div}-\ref{rho_t_div}) for the shell fluids,
and evolution equation for the shell location (\ref{tab_o}) are:
\bea
\dot{C_t} - C_{r}{}' &=& 4\pi\rho\ e^{2B} \delta(f) \frac{dr}{ds}
                    - e^{2B} \ell^2\ \Theta(-f) \nonumber\\
                     &+& 2 (C_r^2-C_t^2) + \frac{4 C_r}{r} + \frac{1-e^{2(B-C)}}{r^2}, \label{Ct_eom}\\
\dot{B_t} - B_{r}{}' &=&-4\pi(\rho+2 P)\ e^{2B} \delta(f) \frac{dr}{ds} \nonumber\\
                     &+& 4\pi\left[\psi_r^2-\psi_t^2+(\xi_r^2-\xi_t^2)\Theta(f)\right]  \nonumber\\
                     &+& C_t^2-C_r^2 - \frac{2 C_r}{r} - \frac{1-e^{2(B-C)}}{r^2},\label{Bt_eom} \\
\dot{\psi_t} - \psi_r' &=& 2 (\psi_t C_r - \psi_t C_t) + \frac{2 \psi_r}{r},\label{psi_eom} \\
\dot{\xi_t} - \xi_r'  &=&  2 (\xi_t C_r - \xi_t C_t) + \frac{2 \xi_r}{r},\label{xi_eom}
\eea
\bea
\ddot{\rho_{g}}
 &=& -\left(\dot{\rho_g}+F(\rho_g+p_g)\right)\cdot\nonumber\\
      & & \left(F(1+\tau_p/\tau_e)-\hat{\dot{B}}+\frac{V\dot{V}}{1-V^2}+
          \frac{\dot{\tau_e}}{\tau_e} \right)\nonumber\\
      &-&F(\dot{\rho_g}+\dot{p_g}) -(\rho_g+p_g)\hat{\dot{F}}
      +\frac{\zeta F^2}{\tau_e} 
       - \frac{e^B\sqrt{1-V^2}}{\tau_e}\cdot\nonumber\\
      & &\bigg[\dot{\rho_g}+F(\rho_g+p_g)+J_U e^B\sqrt{1-V^2} - j\bigg] \\
 &=& -\left(\dot{\rho_g}+\frac{3F\rho_g}{2}\right)
      \left(\frac{3F}{2}-\hat{\dot{B}}+\frac{V\dot{V}}{1-V^2}+
      \frac{\dot{\tau_e}}{\tau_e} \right)\nonumber\\
      &-&\frac{3}{2}(F\dot{\rho_g}+\hat{\dot{F}}\rho_g) +\frac{\zeta F^2}{\tau_e} 
       - \frac{e^B\sqrt{1-V^2}}{\tau_e}\cdot\nonumber\\
      & &\left[\dot{\rho_g}+\frac{3 F\rho_g}{2}+J_U e^B\sqrt{1-V^2} - j\right] 
\label{continuity_g},\\
\dot{\rho_s}&=& -(\rho_s+p_s) F\nonumber\\
            &=& -2 \rho_s F \label{continuity_s},\\
\dot{\rho_\tau}&=& - (\rho_\tau+p_\tau) F - j\nonumber\\ 
            &=& -j, \label{continuity_t}\\
\dot{V}  &=& (1-V^2)\bigg[\frac{2 P (V \tilde{C_t} + \tilde{C_r}+1/r)}{\rho}-(V \tilde{B_t} + \tilde{B_r})\nonumber\\
         &+&\frac{e^B\sqrt{1-V^2} (\ell^2-4\pi[\xi_U^2+\xi_S^2]) }{8\pi\rho}\bigg].\label{V_t_eom}
\eea
In the above, an overdot $(\dot{\ })$ denotes the derivative with respect to $t$ and prime $(')$
the derivative with respect to $r$. These equations are supplemented with ``trivial'' evolution equations for first order gradient variables, i.e. $\dot{z_r}=z_{t}{}'$ and $\dot{z}=z_t$.
The second equality in each of the fluid evolution equations
is after the given equation of state has been substituted in, and for the viscous fluid
a similar relationship for one of the transport coefficients: $\tau_p=\tau_e/2$. 
The variable $j$ denotes
the component of the internal flux $j^a$ in the direction of $u^a$. 
The fluid evolution requires derivatives of the metric intrinsic to the
shell (which in spherical symmetry will only be along the $u^a$ direction); for simplicity we denote such gradients with hatted-dots,
and can be computed with the 4D metric using appropriate combinations of $r,t$ gradients, e.g. $\hat{\dot{B}}=B_t + V B_r$
(and the combination is continuous across the shell despite the individual terms having jumps).
We also introduced
\be
F\equiv \dot{{\bf A}}/{\bf A} = 2 (\hat{\dot{C}} + V/r) \label{F_def}
\ee
representing the fractional change in proper area ${\bf A}(t)$ along the shell.
Note that in first order hydrodynamics, the evolution equation for $\rho_g$ is
a second order PDE (second order ODE in 
spherical symmetry). If all the viscous transport coefficients are zero,
it reduces to the first order, ideal equations (the term in the square brackets
on the last line of (\ref{continuity_g})), and in that case
we directly integrate the latter for $\rho_g$.
Recall that $\rho$ and $P$ in the metric and shell evolution equations are given by (\ref{rho_net}) and (\ref{P_net})
respectively, and here
\bea
\mathcal{A} &=& \tau_e \frac{F(\rho_g+p_g)+\dot{\rho_g}}{e^B \sqrt{1-V^2}}\nonumber\\
            &=&  \tau_e \frac{3F\rho_g/2+\dot{\rho_g}}{e^B \sqrt{1-V^2}}
, \\
\Pi &=& \tau_p\frac{F(\rho_g+p_g-\zeta/\tau_p)+\dot{\rho_g}}{e^B \sqrt{1-V^2}}\nonumber\\
    &=&  \tau_e\frac{F(3\rho_g/2- 2\zeta/\tau_e)+\dot{\rho_g}}{2 e^B \sqrt{1-V^2}},
\eea
where after the second equalities we have again substituted in $p_g=\rho_g/2,\tau_p=\tau_e/2$.

Note that (\ref{Ct_eom}) and (\ref{Bt_eom}) essentially contain the Israel junction conditions,
but directly in terms of our metric variables. I.e., demanding a coordinate system
where the variables are continuous at the shell, but can have discontinuities in gradients,
then it is only the latter terms above that can
balance the delta function terms. These conditions give:
\bea
\Delta C_r &=& - 4\pi\rho e^{B}/\sqrt{1-V^2},\\
\Delta C_t &=& - V \Delta C_r, \\
\Delta B_r &=&   4\pi(\rho+2P) e^{B}/\sqrt{1-V^2}, \\
\Delta B_t &=& - V \Delta B_r \label{delta_B},
\eea
where $\Delta$ refers to the jump in the respective quantity at the shell
(one can be check that the above expressions do coincide with the results computed 
directly using the Israel formalism.)

\subsection{Constraint Equations and Initial Data}

Initial data for the metric evolution is subject to the usual constraint
equations of general relativity. 
The $tt$ component of the Einstein equations can be considered a constraint
equation for $C$ :
\bea
& &C_r'  +  \frac{3}{2} C_r^2 + C_r\left(\frac{3}{r}-B_r\right)-\frac{B_r}{r} \nonumber\\
& &+ 2\pi\left(\psi_r^2+\xi_r^2+
 \frac{2\rho e^{B} \delta(f)}{\sqrt{1-V^2}}\right) 
  +  \frac{1-\ell^2 r^2 e^{2B} - e^{2(B-C)}}{2r^2} \nonumber\\
& &\hspace{0.5in}= \frac{C_t^2}{2}+B_t C_t - 2\pi(\psi_t^2+\xi_t^2).  \label{C_const}
\eea
We have placed time-dependent terms on the right hand side, which for simplicity
we will choose to be zero at the initial time (i.e., a moment of time-symmetry).
The $tr$ component of the Einstein equations can then be considered a constraint
equation for $B$: 
\bea
& &\dot{C_r} + C_t\left(C_r-B_r+\frac{1}{r}\right) - B_t\left(C_r+\frac{1}{r}\right)\nonumber\\
& &+4\pi\left(\xi\Pi -\frac{V\rho e^B \delta(f)}{\sqrt{1-V^2}}\right)=0. \label{e12_eqn}
\eea
Interestingly, at a moment of time symmetry this is trivially satisfied
and $B$ is arbitrary.
At first glance then a simple choice is $B={\rm const.}$ (with an appropriate jump
at the shell location), however then the evolution equation implies
there will be dynamics in $B$, even for a static shell. Instead then,
we will use the evolution equation with all time derivatives 
set to zero to define our choice for $B(r,t=0)$, as then
the static case will be reflected as such in the solution. Specifically, we 
will solve the following for $B(r,t=0)$:
\bea
& &B_{r}{}' -4\pi(\rho+2 P)\ e^{2B} \delta(f) \frac{dr}{ds} + 4\pi(\xi_r^2 + \psi_r^2)\nonumber \\
& & -C_r^2 - \frac{2 C_r}{r} - \frac{1-e^{2(B-C)}}{r^2} = 0.\label{B_id}
\eea
For the shell, a moment of time symmetry requires $V(t=0)=0$,
but the initial position and matter energy densities are arbitrary. For the latter, we will choose
initial conditions to give a static shell when unperturbed, initializing
the matter components following \cite{Danielsson:2017riq}.
For the scalar fields, we set $\psi_t(r,t=0)=0=\xi_t(r,t=0)$,
and freely choose $\psi_r(r,t=0),\xi_r(r,t=0)$, with the particular profiles discussed in Sec.~\ref{BB_id}.

\subsection{Boundary conditions}
For the inner boundary (origin of the AdS region) one can impose regularity through L'Hopital's rule and
requiring $C=C_0(t)+C_2(t) r^2$, and similarly  for $B$ and $\psi$ ($\xi$ does not
extend into the interior). With this, the Einstein equations require $B_0(t)=C_0(t)$,
together with the following conditions at $r=0$:
\bea
\dot{C_t} - 6 C_{r}{}' + B_r{}' &=& 
                    -e^{2 B} \ell^2 - 2 C_t^2  , \label{Ct_eom_orig}\\
\dot{B_t} - 2 B_{r}{}'+ 3  C_{r}{}' &=&
                    -4\pi \psi_t^2
                    +C_t^2 ,\label{Bt_eom_orig} \\
\dot{\psi_t} - 3 \psi_r' &=&  -2 \psi_t C_t .\label{PI_eom_orig}
\eea

For the outer boundary (in the AF region), one can use maximally dissipative boundary conditions (e.g.~\cite{Calabrese:2003vx}). 
For instance, for the scalar field $\Psi$, its equation of motion when written in first order form 
(with $\Pi\equiv \Psi_{,t}$, $\Phi=\Psi_{,r}$) is given by
\bea
\dot{\Pi} &=& \Phi' + R_{\Pi},   \\
\dot{\Phi} &=& \Pi'+ R_{\Phi},
\eea
with $R_{\Pi}, R_{\Phi}$ the remaining terms of the corresponding equations not belonging to the principal part.
The incoming (outgoing mode) at $r=R_{out}$ is $\Pi + \Phi$ ($\Pi-\Phi$). Maximally dissipative boundary conditions
define incoming mode(s) as related to (and bounded by) the outgoing ones. 
For simplicity we can do this at the level of the time derivatives of the modes; that is,
\bea
\dot{\Pi} + \dot{\Phi} &=& a \left(  -(\Pi' - \Phi') + R_{\Pi} -  R_{\Phi} \right ),\label{md_eqn1}\\
\dot{\Pi} - \dot{\Phi} &=& -(\Pi' - \Phi') + R_{\Pi} -  R_{\Phi} \label{md_eqn2}.
\eea
The first line states that the incoming mode is proportional (with proportionality constant $a$) to the outgoing mode. 
If $|a|<1$, the condition is said to be maximally dissipative, with $a=0$ describing purely outgoing modes. 
The special case $|a|=1$ corresponds to the reflecting case.

Now, solving for the time derivatives in (\ref{md_eqn1}-\ref{md_eqn2})
we derive what we should impose at the outer boundary point $r=R_{out}$:
\bea
\dot{\Pi}  &=& \frac{(a+1)}{2} \left( -(\Pi' - \Phi') + R_{\Pi} -  R_{\Phi}\right),   \\
\dot{\Phi} &=& \frac{(1-a)}{2} \left( (\Pi' - \Phi') - R_{\Pi} +  R_{\Phi}\right).
\eea
In the code, we implement the above with $a=0$ for the scalar field. For simplicity we do the same with the metric variables, 
as they also obey wave equations. However, such maximally dissipative conditions are not fully consistent with the constraints,
introducing an error that scales as $1/R_{out}$. To mitigate this problem, as described in Appendix~\ref{map_rx_A},
we control the mapping between radial and code coordinates to push the outer boundary to be out of causal contact
with the bubble for the duration of a given simulation. 


\subsection{Simplified Shell Dynamics with an External Source}\label{diss_model}

As discussed above, one of our goals motivating the particular choice
of metric, coordinate conditions, etc., is to have a scheme that could eventually
be generalized beyond spherical symmetry. Also, before taking on such an endeavor,
we want to get some indication on how a gravitational wave might interact
with the bubble, using a scalar field that can freely propagate across the bubble
as a proxy for a gravitational wave. 
However, even in spherical symmetry, with these particular choices
there are various complications that arise, discussed in the Sec.~\ref{diss_EKG_results},
that make it challenging to extract useful results on the physics of black bubbles
in the large mass limit of interest.
Here then we introduce a simplified model
that allows us to explore black bubble stability in this regime, but it is only
applicable to spherically symmetric systems, and cannot model interaction
with a bulk scalar field.

For this simplified model of the black bubble,
consider a shell enclosing an AdS spacetime, with a Schwarzschild exterior that
can contain unspecified matter.
Here, we parameterize all shell quantities with proper time $\tau$ on the shell.
We will model interaction with the exterior matter via a flux function $J_U(\tau)$.
We also only consider a vacuum
AdS interior, though in the linear analysis in Appendix~\ref{vac_sd} we will allow a small
internal mass to model some prior interaction that led to interior energy.

Thanks to spherical symmetry we can solve such a model 
by integrating the Einstein and shell fluid equations purely at the shell location,
making sure we 
self-consistently incorporate the back-reaction of the external
flux on the gravitational mass of the bubble. A straight-forward way
to compute the latter is to assume the external flux is coming from the scalar field
$\xi$ as in the full spacetime model, and imposing the Einstein equations
for this system just exterior to the shell, then taking the limit onto the shell (again, keeping the interior spacetime fixed). 
The effect of the scalar field can then simply be modeled as some
freely specifiable $J_U(\tau)=-\xi_S(\tau)\xi_U(\tau)$; i.e.
we do not need to know what particular external scalar field 
profile would be needed to lead to such a flux at the shell.

We impose the following ansatz for the spacetime:
\be
ds^2 = -g(t,r) dt^2 + \frac{dr^2}{f(r,t)} + r^2 d\Omega^2.
\ee
Exterior to the shell we set $f(r,t)= 1 - 2 m(t,r)/r$, so $m(t,r=R)$ will represent
the exterior Schwarzschild mass of the spacetime. We still 
have coordinate freedom with this ansatz to rescale $t$
by an arbitrary function of itself, and do so to impose $g(t,r=R)=f(t,r=R)$,
i.e. evaluated at the shell the exterior metric looks exactly like the Schwarzschild
solution but with a time-dependent mass. 
Interior to the shell we use the following static form for the AdS spacetime:
$f(r)=g(r)=1+r^2\ell^2/3$. Below, metric quantities that are discontinuous
across the shell are labeled with a subscript $L$ when evaluated
just to the left (interior) of the shell, 
and with a subscript $R$ just to the right (exterior) of the shell.

In this section we will use the over-dot to denote change with respect
to proper time, e.g. $\dot f \equiv d f(\tau)/d\tau$. With that notation,
the evolution equations for the shell, its internal
energy components, and $m(\tau)$ are
\bea
\dot{R}&=&V, \label{Rdot} \\
\dot{V}&=&Q_L Q_R\bigg[\frac{2P}{\rho R} + \frac{1}{Q_L+Q_R}
        \bigg(\ell^2\left[\frac{1}{4\pi\rho}-\frac{R}{2 Q_L}\right]\nonumber\\
        & &\hspace{0.5in}-(\xi_S^2+\xi_U^2)\left[\frac{2\pi R}{Q_R}+\frac{1}{\rho}\right] \bigg)\bigg] \nonumber\\
        & &+\frac{Q_L Q_R - 1 - V^2}{2R}\label{V_tau_eom_J1},\\
       &=&  \frac{R \ell^2 Q_R - 4\pi Q_L \left[R(\xi_S^2+\xi_U^2)-4 P Q_R\right]}{2(Q_L-Q_R)} \nonumber\\
        & &+\frac{Q_L Q_R-1-V^2}{2R}, \label{V_tau_eom_J2}\\
\ddot{\rho_g} &=& - \frac{3 \rho_g (R \dot{V} + V^2)}{R^2} - \frac{5V}{R} \dot{\rho_g} 
                    - \frac{\tau_p}{\tau_e}\left[\frac{2V}{R^2}(3V\rho_g + \dot{\rho_g} R) \right] \nonumber \\
                &-&\hspace{-0.025in}\frac{1}{\tau_e}
                 \bigg[(\dot{\tau_e}+1)\left(\frac{3V}{R}\rho_g+\dot{\rho_g}\right) 
                - \frac{4\zeta V^2}{R^2} -j -\xi_U\xi_S\bigg], \label{eps_pp_J} \\
\dot{\rho_s}&=&-\frac{4 V \rho_s }{R}, \label{RHOSdot} \\
\dot{\rho_\tau}&=&-j, \label{RHOTAUdot} \\
\dot{m}&=&\frac{4\pi R^2 Q_R (Q_R \xi_U - V \xi_S)(Q_R \xi_S- V \xi_U)}{f_R}, \label{Mdot}
\eea
where $f_R\equiv{1-2m(\tau)/R(\tau)},\  f_L\equiv1+R(\tau)^2\ell^2/3,\ Q_R\equiv\sqrt{f_R+V^2}$, $Q_L\equiv\sqrt{f_L+V^2}$,
and 
\bea
P&=&p_g + p_\tau + p_s +\tau_p\left[\dot{\rho_g}+\frac{2V(\rho_g+p_g)}{R}\right]\nonumber\\ 
 & &- \frac{2 V \zeta}{R},\label{net_P}\\
\rho&=&\rho_g+\rho_\tau+\rho_s + \tau_e\left[\dot{\rho_g}+\frac{2V(\rho_g+p_g)}{R}\right]\label{net_rho}.
\eea
The first equation for $\dot{V}$ (\ref{V_tau_eom_J1}) stems from (\ref{tab_o}), and for 
reference below that in (\ref{V_tau_eom_J2}) we also include a form coming directly  
from the junction condition proportional
to the net pressure (or equivalently eliminating $\rho$ from
the previous equation using the junction condition proportional to $\rho$).
Again, the external source functions $\xi_U(\tau)$ and $\xi_S(\tau)$ can be considered 
freely specifiable; setting $\xi_U(\tau)=\xi_S(\tau)$ models the perfectly absorbing conditions.

For reference, as this will be needed for the flux $j$ as described in Sec.~\ref{AdSBubble}, 
the exterior proper acceleration is
\be\label{aR_full}
a_R = \frac{4\pi R^2(\xi_U^2+\xi_S^2)+2\dot{V}R +1 - f_R }{2 Q_R R},
\ee
where gradients of $f$ and $g$ appearing in its definition (\ref{four_a_def})
have been eliminated using the Einstein equations.

\subsection{Simplified dissipation}
In the above equation for $\rho_g$ (\ref{eps_pp_J}) we have included all the three
relevant transport coefficients, $\tau_e, \tau_p$ and $\zeta$, which in general
are all dependent on $\rho_g$ (hence $\tau$), and need to be non-zero to give a well-defined,
hyperbolic theory. However, experimentation suggested $\tau_e, \tau_p$ have
little effect on the dynamics of the bubble. This can be understood 
by rewriting (\ref{eps_pp_J}) as follows. Let
$\mathcal{I}\equiv 2V(\rho_g+p_g)/R + \dot{\rho_g}$; i.e. $\mathcal{I}=0$
is just the flux-free perfect fluid equation of motion. Then in terms of
$\mathcal{I}$, (\ref{eps_pp_J}) becomes
\be
\dot{\mathcal{I}} = - \mathcal{I} \left[\frac{\dot{\tau_e}+1}{\tau_e} +\frac{2V}{R}\left(1+\frac{\tau_p}{\tau_e}\right)\right]
+ \frac{4\zeta V^2}{\tau_e R^2} + \frac{j + \xi_U\xi_S}{\tau_e}.
\ee
This suggests that in spherical symmetry the parameters $\tau_e,\tau_p$ essentially only control return to hydrodynamic evolution
when starting from beyond-ideal conditions; i.e. ignoring the fluxes, if $\zeta=0$ as with a conformal fluid, and we
begin in equilibrium where $\mathcal{I}=0$, then $\mathcal{I}$ will remain zero for all time, and
$\rho_g$ will always behave like an ideal fluid. 
The $\zeta$ term, being proportional
to $V^2$, becomes important with nonlinear perturbations, and since it is always
positive it is consistent with the intuition that this must come from 
dissipation removing kinetic energy from the motion of the bubble and depositing it in the gas.
This is likewise consistent with the equation of motion for the bubble
(\ref{V_tau_eom_J1}) : if $\mathcal{I}=0$, the $\tau_e,\tau_p$ terms drop
out from the expressions for the net pressure and energy density (\ref{net_P}-\ref{net_rho}),
and $\zeta$ controls the damping of the shell 
\be
\dot{V}\approx - V \frac{4\zeta Q_L Q_R}{R^2 \rho} + ...,
\ee
where $...$ denote terms that do not depend on any of the dissipation parameters.

Motivated by these observations, we set $\tau_e=\tau_P=0$ (starting from
(\ref{eps_pp_J}) one needs to first multiply by $\tau_e$, then take
the limit). 
With that, (\ref{eps_pp_J}) becomes 
\be
\dot{\rho_g} = -\frac{3V}{R}\rho_g + \frac{4\zeta V^2}{R^2} + j + \xi_U\xi_S, \label{RHOGdot} \\
\ee
with
\bea
P&=&\frac{\rho_g}{2} - \rho_\tau + \rho_s  - \frac{2 V \zeta}{R},\\
\rho&=&\rho_g+\rho_\tau+\rho_s.
\eea
Note that it would be trivial to add $\zeta$-dissipation to the other
shell components (with their sum then appearing in the expression
for $P$ above), or to split the energy flux $\xi_U\xi_S$ in some prescribed
manner to the other matter components.

\section{AdS Black Bubble Matter}\label{AdSBubble}

In the preceding sections we have described all of the components of the black bubble
we study here, with the exception of the key property essential for it to
be an astrophysically viable compact object candidate: stability. This, in principle,
is achieved via an appropriate choice of internal flux $j$ between
the gas and brane components. We begin by reviewing the original suggestion
for this given in~\cite{Danielsson:2017riq}, then describe its short coming 
and novel suggestions to overcome it.

As already mentioned, the shell is composed of three constituents,
a brane with EOS $p_\tau=-\rho_\tau$, a gas of massless particles with EOS $p_g=\rho_g/2$, and a stiff fluid with EOS $p_s=\rho_s$, 
which are required based on physical and kinematic grounds. Let us review how this can be motivated from string theory.
Inside of the shell there is an AdS space with a negative cosmological constant. The main idea behind this scenario is that space 
time is unstable against decay to an AdS space. Usually, such a decay is heavily suppressed, but when matter threatens to collapse 
and form a black hole, the nucleation is enhanced for entropic reasons. If a bubble forms, the infalling matter can turn into 
massless open strings, attached to the shell, carrying an entropy close to the one carried by a genuine black hole. This is similar 
to what is argued to happen in the case of fuzz balls. From string theory, it is expected that the scales associated with the 
negative cosmological constant, as well as the tension of the brane, are high energy. Certainly beyond what is presently accessible 
through accelerator experiments and possibly close to the Planck scale. 

The positive energy of the brane is supposed to closely match the negative energy of the vacuum inside of the shell. The mass of 
the system is then carried by the matter on top of the shell. If the shell has a radius given by $\frac{9R_s}{8}$, where $R_s$ is
the Schwarzschild radius, then the
Israel-Darmois junction conditions forces matter to have the equation of state of a gas of massless particles. This special 
radius is often referred to as the Buchdahl radius. Such a matter component, composed of massless open strings attached to the 
brane, is natural from a string theoretical point of view. In order for the gas to be able to carry an entropy comparable to the 
one of a black hole, the number of degrees of freedom needs to be large. This can be accomplished if the endpoints of the strings 
are supported, not by the 2+1 dimensional brane itself, but by a huge number of lower dimensional branes dissolved in it. The need 
for such dissolved branes can also be seen by examining the junction conditions. This is where the stiff gas enters.

In string theory, 4D supersymmetric black holes can be constructed using 3-branes wrapping internal 3-cycles. Such branes will be point like from the 
4D space time point of view. As suggested  in~\cite{Danielsson:2017riq}, black bubbles in 4D can be obtained as 3-branes 
polarized into a 5-brane, still wrapping the internal 3-cycles. This 5-brane can still carry 3-brane charges represented by 
magnetic fluxes inside of the 5-brane. Ignoring the internal three dimensions, this is captured by the DBI action given by:
\be
S= \int d^3 \sigma T_2 \sqrt{ -{\rm det} (h_{\mu \nu} +\cal{F}_{\mu \nu} )}  ,
\ee
where $T_2=\rho_ \tau$ is the tension of the shell, and $\cal{F}_{\mu \nu}$ is the magnetic flux inside of the brane. The flux is 
quantized, and the energy density is schematically given by $4\pi T_2 \sqrt{r^4+N^2}$, where $N$ is an integer counting the number of 
dissolved branes. Note that if we formally take the radius of the shell to zero, the contribution of the shell goes away and the energy
is dominated by the mass of the D-particles. For a large shell, the contribution from the magnetic flux will be suppressed and, 
as explained in~\cite{Danielsson:2017riq}, have an energy density of order $N^2/r^4$ with the equation of state of a stiff gas.  
On top of this, there are massless fluctuations of the gauge fields. The number of such modes is order $N^2$ and they give 
rise to the $\rho_g$ that will carry the entropy. 

In this way, one can solve the junction conditions, at the Buchdahl radius, using components motivated from string theory. 
For this setup to be a viable alternative to an ordinary black hole, it is not enough to find a critical point, it must also 
be stable. Unfortunately, this is not the case unless there is nontrivial dynamics involving energy exchange between the various 
components. The challenge is to find out what kind of dynamics is necessary, and whether this is what to expect from string theory. An argument for how stability could be obtained, based on thermalization at the local Unruh temperature, was 
given in~\cite{Danielsson:2017riq}. Let us elaborate a bit on the argument presented there. 

The shell will be heated through a non-zero Unruh temperature from the outside due to its non-zero proper acceleration sitting 
at a constant radius in the Schwarzschild metric. \footnote{Note that if the shell were brought towards the horizon, the 
Unruh temperature would increase towards infinity. As observed from infinity the temperature will, when the redshift is 
taken into account, approach the Hawking temperature $T_H$. The temperature of the Buchdahl shell will be slightly lower and given 
by $\frac{64}{81}T_H$.}   (There will be no such heating from the inside AdS region since there is a threshold for the 
acceleration\cite{Deser:1997ri}). If the temperature of the shell is a bit lower than the Unruh temperature, the gas will 
absorb Unruh quanta. Each mode will act as a little antenna. Thus, the shell will absorb at a rate of $N^2 \times R^2 \times T^4$. 
Since $N \sim R$ and $T \sim 1/R$, the total power of absorption will be of order one. That is, the gas can absorb a mass of 
order $M$ in light crossing time $R$. This suggests a term $\frac{\dot{T}}{T} \rho_g$, with no further suppression, 
contributing to the source term $j$. The Unruh quanta are not real, so energy needs to be supplied from the system itself for 
them to be created. In our model, it is the tension of the brane that is reduced in order to power the increased energy density 
of the gas. Note that the probability for energy to radiate off the system into the surrounding space, reducing the total energy, 
is heavily suppressed. Heuristically, the rate would not be order one but reduced by a factor $1/N^2$ due to self-absorption into 
the other modes.  The resulting loss of energy is therefore of the same order as Hawking radiation and can be ignored in our 
analysis. The fact that the large number of degrees of freedom make it so entropically favorable for energy to get stuck to the 
brane, is the reason why the system can so closely mimic a black hole; i.e. appearing to external
observers as a near perfect black body of similar size and temperature to that of the equivalent mass black hole.

When the area of the shell changes, the number of dissolved branes, $N$, needs to change. Their energy are subleading, but 
when $N$ changes one would expect that the massless perturbations of the gauge field need to change too. These carry important 
amounts of energy, and therefore one expects a contribution of the form $F \rho_g$ to $j$.

We have thus argued, from a microscopic point of view, for the presence of the two terms in our ansatz for $j$ : one proportional 
to changes in the temperature $T$, the other to changes in the area $F$. In the specific model described next, these terms are parameterized by 
constants $\alpha$ and $\beta$ respectively. In ~\cite{Danielsson:2017riq} values for $\alpha$ and $\beta$ consistent with a 
quasi-static approximation were considered. However, such an approximation is not relevant for any real physical process where 
the shell is perturbed by infalling matter. In the discussion that follows, we will perform a more careful analysis constraining the parameters so 
that we obtain a self stabilizing shell. We will also verify the results using numerical methods. Interestingly, the constraint 
we find has a very simple and suggestive form.

\subsection{Specific flux model}

The total energy density $\rho$ and pressure $P$ sourcing Einstein's equations at the bubble location are the 
sum of the distributional matter terms
\bea
\rho&=&\rho_g+\rho_s+\rho_\tau,\\
P&=&p_g+p_s+p_\tau = \frac{1}{2}\rho_g+\rho_s-\rho_\tau,
\eea
where here we ignore any viscous corrections $\mathcal{A}$ and $\Pi$ to these quantities. As mentioned,
we will require that
the gas has a thermal component at the instantaneous local Unruh 
temperature of an observer on, but {\em outside} the shell:
\be
T = \frac{a_{_R}}{2\pi}, \label{TU}
\ee
where the subscript $()_{_R}$ denotes the quantity is evaluated to the
right (outside) of the shell. The vectors $u^a$ and $s^a$ are the same vectors
on either side of the shell, as are their coordinate representations
in our coordinate system, however their gradients orthogonal to the shell
are generally discontinuous across it; in particular, the magnitude
of the 4-acceleration evaluated using the EKGH metric is
\be\label{four_a}
a \equiv \nabla_a u^b u^a s_b = \frac{B_r + V B_t + \dot{V}/(1-V^2)}{e^B\sqrt{1-V^2}},
\ee
and from (\ref{delta_B}) one can see how $a$ will jump across the shell.

The continuity equation (\ref{continuity}) is only required to be satisfied
by the net fluid quantities $\rho$ and $P$, and it is up to us to specify any internal
interactions between the fluid constituents. As discussed in the previous
section, the brane will provide the energy for heating/cooling,
and any response to changes in the area of the shell. Since the stiff fluid
component is subleading, we only consider a flux $j$ between the brane
and gas, leading to the individual continuity equations
given in (\ref{continuity_g}-\ref{continuity_t}) and (\ref{eps_pp_J}-\ref{RHOTAUdot}),
which we repeat here for convenience (without dissipative terms):
\bea
\dot{\rho_g} &=& - (\rho_g + p_g) F + j = -\frac{3}{2} \rho_g F + j \label{cont_g},\\
\dot{\rho_\tau} &=& - (\rho_\tau + p_\tau) F -j = -j, \\
\dot{\rho_s} &=& - (\rho_s + p_s) F = -2 \rho_s F, 
\eea
and recall $F$ represents the fractional change in proper area along the shell 
trajectory (\ref{F_def}).

\subsubsection{Internal energy exchange and stability}
To obtain guidance leading to a concrete prescription for the internal flux, we begin
by assuming the gas component $\rho_g$ is purely thermal, namely
\bea
\rho_g &\propto& N^2 T^3,\label{rho_g_thermal}\\
p_{g} &=& \rho_g/2,
\eea
where again $N$ is the number of particles. 
With the assumption that $N$ is fixed
\be
\dot{\rho_g} = 3 \rho_g \dot{T}/T.\label{T_dot}
\ee
The continuity equation (\ref{cont_g}) 
gives an evolution equation for $\rho_g$; therefore if there was no source $j$ 
then (\ref{T_dot}) would simply tell us how the temperature evolves.
On the other hand, as discussed above, it is assumed that locally the brane can interact
with the gas on timescales much smaller than any macroscopic dynamical
timescale to always keep the temperature 
fixed at the Unruh temperature (\ref{TU}). In that
case the continuity equation can be viewed as the definition of the
flux of energy $j$ coming from the brane required to enforce this; i.e. we want
\be
\dot{a_{_R}}/a_{_{R}}=\dot{T}/T, 
\ee
which requires the flux to be
\be\label{flux_eqn}
j \equiv 3\rho_g \left(\dot{a_{_R}}/a_{_{R}} + F/2\right). 
\ee 
The appearance of the term $F$, representing the fractional change in area
as the shell moves (\ref{F_def}), exactly cancels the ``usual'' response
of energy density to such a change in area (\ref{cont_g}). This comes
from us assuming that the internal interaction in the shell is entirely
driven by changes in the local proper acceleration, and moreover that
the interaction forces (\ref{rho_g_thermal}) to always be satisfied.
The quasi-stationary analysis given in~\cite{Danielsson:2017riq} suggested
this was adequate for stability of the black bubble. However as we
show in appendices~\ref{dynamicalproperacc} and~\ref{vac_sd},
~\cite{Danielsson:2017riq} ignored a dynamical component to changes in the 4-acceleration
that has a destabilizing effect. Motivated by this observation, and the string theory
considerations discussed above,
we propose the following
modification of (\ref{flux_eqn}) to model a broader class of internal interaction
\be\label{flux_eqn_G}
j \equiv 3\rho_g \left(\alpha\ \dot{a_{_R}}/a_{_{R}} + \beta\ F/2\right).
\ee 
Here, $\alpha$ is a constant controlling changes to the internal state
of the shell in response to changes in the Unruh temperature,
while $\beta$ is a constant controlling corresponding changes
when the material compresses ($F<0$) or expands ($F>0$). This model is clearly
ad-hoc, though at least can be used to illustrate what kind of internal
flux may be needed to stabilize the black bubbles, and serve
as a guidepost for future investigation of bubble constructions within
a self-consistent theory.

\subsubsection{Alternative flux model}
We can also consider the gas temperature does not instantaneously adjust to the local Unruh temperature
$T_u=a_R/2\pi$, but instead relaxes to it on a characteristic timescale $\tau_u$ via
\be\label{T_eqn_def}
\dot T = \frac{1}{\tau_u} \left(\frac{a_{_{R}}}{2\pi} - T\right).
\ee
Carrying this through a similar calculation as above, and
again generalizing with parameters $\alpha$ and $\beta$, defines an alternative flux
option given by:
\be\label{j_u_def}
j \equiv 3\rho_g \left(\frac{\alpha}{\tau_u}\left(\frac{a_{_{R}}}{2\pi T} - 1\right) + \beta\ F/2\right).
\ee
With this prescription for the flux, $T$ is evolved as an independent variable.


\section{Implementation specifics}\label{implementation}
With the goal of studying the dynamical behavior of the AdS black bubble
and potential observable consequences, we wrote two different codes
for an efficient exploration. These implement the EKGH system in Sec.~\ref{EKG_numerics}
which we employ to assess the full spacetime dynamics, and the shell model in Sec.~\ref{model_numerics}
to efficiently scrutinize the bubble's behavior.

\subsection{Einstein-Klein-Gordon-Hydrodynamic system}\label{EKG_numerics} 
For the most part, our discretization and solution of the 
EKGH system outlined in Sec.~\ref{EKG_formalism} is straight-forward
and follows standard finite difference techniques.
Specially, for the PDEs away from the shell location we use second order accurate stencils for spatial gradients,
add Kreiss-Oliger style dissipation~\cite{Kreiss_1973}, and for the time integration use a second order accurate explicit Runge-Kutta (method of lines) scheme. 

Special treatment is needed at the
location of the shell, where even with our choice of a continuous metric across it,
there are discontinuities in gradients there, hence finite difference methods are not applicable.
As discussed before, in spherical symmetry, where there are no propagating gravitational wave degrees of freedom,
one can treat the shell location as a ``boundary'' of both the interior
and exterior spacetime, connecting them via the Israel junction conditions. However,
this is not possible in general, as the shell location is not a boundary of the spacetime, and gravitational
waves can freely propagate across it. We therefore want to implement a scheme
that can integrate the field equations self-consistently across singular surface layers.
Here we do so via a weak-form, finite volumed inspired strategy, described in Sec.~\ref{sec_weak}.
In spherical symmetry in our chosen coordinates this allows the gauge waves present
in the metric variables $B$ and $C$ to freely propagate across the shell location,
as well as our $\psi$ scalar field proxy for gravitational waves, without imposing
any boundary conditions. For simplicity, we have only implemented this to first order
accuracy at present, hence even though everywhere else the discretization is second
order accurate, we only expect global first order convergence in the continuum limit.

It would be complicated to perform these weak-form integrations
over a layer that moved on the coordinate grid. Therefore, as described in Sec.~\ref{rx_map},
we define a separate
spatial code coordinate $x$, and dynamically adjust the mapping to the metric
coordinate $r$ so that the bubble location is always at a fixed $x$ coordinate.
Of course this is easy to do in spherical symmetry, and one might worry
that generalizing this would be very challenging. However, we note that
much more sophisticated ``dual frame'' schemes have already been
successfully implemented in binary black hole merger simulations in full $3+1D$~\cite{Scheel:2006gg}
(see also ~\cite{Hilditch:2015qea}).
There, the black hole excision surfaces are kept at fixed code locations,
and it should be possible to adapt those techniques to bubble spacetimes, at least
prior to any bubble collisions.

Note that in spherical symmetry one can also solve the
constraint equations in lieu of one or both of the evolution equations
during evolution, as effectively the scalar field drives all the non-trivial dynamics then.
Empirically we have found solving (\ref{e12_eqn}) for $C$ instead of the evolution
equation (\ref{Ct_eom}) makes it easier to achieve stable evolution near the origin.
Solving constraints instead of evolution equations is {\em not} easy to generalize
to spacetimes without any symmetry; however, here the origin difficulties are entirely
because of spherical symmetry, and would not be present in, for example, a Cartesian
based coordinate system.

In Sec.~\ref{BB_id} we list particular initial conditions we use for
the shell matter and scalar fields.

\subsubsection{Weak form integration}\label{sec_weak}

Here we outline the idea behind a weak form integration, leaving the description
of the particular stencil used in our implementation in the code to Appendix~\ref{sec_wfs}.

Equations (\ref{Ct_eom})-(\ref{xi_eom}) are all quasi-linear wave equations of the form
\be\label{qle}
\dot{f}(t,r)-g'(t,r)+h(t,r)+\delta(r-R) S(t,r) = 0,
\ee
as would the full 3+1D Einstein equations in harmonic form be.
As mentioned, we discretize this using standard finite difference
methods everywhere except at the shell. At that surface, here the point $r=R$
(which for now we consider to be constant),
we apply the following finite volume, weak-form discretization.
First, multiply the equation by a test function $v(r)$ that only has support within
a cell of width $2\Delta r$ about the shell ($v(r)=0$ for $|r-R|\geq\Delta r$), and integrate over 
the spatial volume of the cell:
\be\label{svi}
\int (\dot{f} - g' + h +\delta(r-R) S) v dr = 0.
\ee
For simplicity let $v(R)=1$, and integrate the gradient term by parts, 
$g'v = (gv)' - g v'$, giving
\be\label{iqle}
\int ([\dot{f} + h] v + g v') dr  = - S(t,R).
\ee
This is an improvement to before, both because we have been able to evaluate the delta function,
and we have shuffled the spatial gradient from $g$ to $v$, the former which has a step at $r=R$ (as it must
so that its gradient can compensate for the delta function in the equation of motion).
I.e., we are free to choose  $v(r)$ to be sufficiently regular
so that $v'$ is finite within the cell, hence $g v'$ is well defined and simple to evaluate,
whereas before $g'v$ was not.

If the shell moves, i.e. $R=R(t)$, the above equation becomes more complicated to
regulate, since the 
time derivative $\dot f (t,r)$ in (\ref{qle}) is the partial of $f(t,r)$
with respect to $t$ at constant $r$, not constant $R$. Hence, in a typical
wave equation where $g$ and $f$ are related, even if there is no
singular behavior in time variation tangent to the shell,
discontinuities in gradients orthogonal to the shell get spread
into both $(\dot{})$ and $(')$ discontinuities, as the $t$ and $r$ coordinates
are not aligned with the $\tau$ and $s$ coordinates tangent and orthogonal
to the shell, respectively. There are several conceivable ways to deal with such
a situation. One is to extend (\ref{svi}) to an integration over a space-time volume.
Another is to choose coordinates that reduce to $(\tau,s)$ along the world
line of the shell. A third, that we have chosen to use, described in 
the next section, and detailing its consequences for the weak-form
integration in Appendix~\ref{sec_wfs}, is to introduce a map $x(t) \leftrightarrow r(t)$
between the metric $r$ and code $x$ coordinates such that the shell is always at a constant
$x$, and then perform the spatial integration (\ref{svi}) over a cell of width $2\Delta x$.

\subsubsection{Mapping between radial metric and code coordinates}\label{rx_map}

We represent the various fields in our EKGH system on a uniform
mesh in a coordinate $x\in [0..x_{out}]$, with the following key properties :
\begin{itemize}
\item $x(r=R(t),t)=R(0)\equiv x_0$ (the shell stays at a constant $x=x_0$)
\item $x(r=0,t)=0$ ($x=0$ maps to $r=0$)
\item $x(r=R_{out},t)=x_{out}$ (the outer boundary is at a fixed $r$ and $x$)
\item $\partial x(r,t)/\partial r|_{r=R(t)}=1$ (the map is at least once-differentiable at the shell location, and $dx$ and $dr$ have the same scale there).
\end{itemize}
We use polynomial functions for the map; the particular expressions are not too enlightening,
so we list them in Appendix~\ref{map_rx_A}. Note that this is not a coordinate
transformation: we still evolve the metric functions $B$ and $C$ (\ref{metric}) and their partials
$B_r,C_r$ and $B_t,C_t$ with respect to $r$ and $t$ respectively. Another way
then to think of this map is as a non-uniform, time-dependent discretization
of $r$. The map will break down if the shell moves too far from its initial position,
though this is only a problem for unstable bubbles.

\subsubsection{Initial data}\label{BB_id}
Our typical initial conditions consist of a static black bubble enclosing empty AdS
spacetime, and then some prescribed external pulse for either of $\psi(r,t=0)$
or $\xi(r,t=0)$ (with $\psi_t(r,t=0)=\xi_t(r,t)=0$) that will subsequently interact
with the shell to perturb it (for unstable bubbles numerical truncation error by itself
will ``perturb'' the shell, causing it to either accelerate outward or collapse to a black hole,
but this is not controllable in that the ``perturbation'' converges away with resolution).
Specifically, given a desired initial $R_0=9 m_0/4$ for the bubble, we set the shell components
following\cite{Danielsson:2017riq} as \footnote{Note that their analysis only gives
a unique decomposition in the large mass (radius) limit, and there are several conceivable ways
of extrapolating that to $m=0$; equations (\ref{rho_g_id}-\ref{rho_t_id}) is one particular
possibility.}:
\bea
& &\rho_g(t=0)  =  \frac{\ell R_0 + (\ell R_0 - \sqrt{3}) \sqrt{1+ \ell^2 R_0^2/3}}{12\pi\ell R_0^2\sqrt{1+ \ell^2 R_0^2/3}},\label{rho_g_id}\\
& &\rho_s(t=0)  =  \frac{\sqrt{3}}{16\pi\ell R_0^2}\label{rho_s_id},\\
& &\rho_\tau(t=0)  = \nonumber\\
& &\ \ \  \frac{4\ell^3 R_0^3 + 8\ell R_0 + (\sqrt{3}-8\ell R_0)\sqrt{1+ \ell^2 R_0^2/3}} 
                        {48\pi\ell R_0^2\sqrt{1+ \ell^2 R_0^2/3}}.
\label{rho_t_id}
\eea
We set 
\bea
\xi(r,t=0)&=&\frac{A_\xi}{(\Delta_\xi)^8} \left(r-(R_\xi-\Delta_\xi)\right)^4 
                                       \left(r-(R_\xi+\Delta_\xi)\right)^4,\nonumber\\
          & &R_\xi-\Delta_\xi < r < R_\xi+\Delta_\xi, \label{sf_id} \\
          &=&0 \ \ \ {\rm elsewhere},
\eea
where $A_\xi,R_\xi,\Delta_\xi$ are constants, and similarly for $\psi(r,t=0)$.

\subsection{Simplified Shell model}\label{model_numerics}

The ODE equations governing the shell model (\ref{Rdot},\ref{V_tau_eom_J2},\ref{RHOSdot},\ref{RHOTAUdot},\ref{Mdot},\ref{RHOGdot}) can be integrated straightforwardly with 
the flux $j$ (\ref{flux_eqn_G}) for the instantaneous adjustment to the Unruh's temperature
of the gas. If, on the other hand, we employ the alternative flux prescription,
we augment the evolution equations with (\ref{T_eqn_def}) and the flux given instead by (\ref{j_u_def}).
The resulting equations are integrated with a standard fourth order Runge Kutta scheme.
Initial data is given by the static black bubble described in section \ref{BB_id} and we consider
its interaction with a perturbation given by $\xi_S(\tau),\xi_U(\tau)$. 
We define these sources via superposition of functions of the form
\begin{equation}\label{shell_pert}
\xi_S(\tau)=\xi_U(\tau)=A_{\xi} \left ( e^{-((\tau-\tau^a_{\xi})/\sigma_{\xi})^2} + e^{-((\tau-\tau^b_{\xi})/\sigma_{\xi})^2}\right ).
\end{equation}
Setting $\xi_S=\xi_U$ corresponds to the maximum rate of absorption of energy by the gas (\ref{RHOGdot}).
Finally, as we employ this code to explore the large $m$ regime, given the disparate length scales involved (bubble mass,
perturbation value and timescale of interest) we adopt quadruple precision.


\section{Applications/dynamics}\label{results}

To explore the stability
of black bubbles in the large mass limit, we use the simplified model
described in Sec.~\ref{diss_model} and \ref{model_numerics}. These results are presented
in Sec.~\ref{diss_model_results}. In Sec.~\ref{diss_EKG_results}
we show some results from the full model described in Sec~\ref{EKG_formalism} and \ref{EKG_numerics},
focusing on issues that would need to be overcome going beyond spherical
symmetry, and results from scalar field evolution on a fixed bubble background.

\subsection{Numerical results from the shell model}\label{diss_model_results}

We now focus on the simplified model described in Sec.~\ref{diss_model} and investigate a couple of interesting cases
with parameters guided by a linear stability analysis of the system (Appendix~\ref{vac_sd}). 

We impart a perturbation of the form (\ref{shell_pert}) to the shell which effectively
imply ``hitting'' it twice : 
the first at $\tau=\tau^a_{\xi}$ 
to take it away from the static solution, and a second one at $\tau=\tau^b_{\xi}= 15 \tau^a_{\xi}$ to further perturb 
the intermediate state before it achieves equilibrium (if stable).
For each perturbation we evolve with two choices for the parameters $\{\alpha, \beta\}$.
The first (Case A) uses the constants $\alpha=0.4$ and $\beta=0.1$. As we show,
this yields stable bubbles, but their final equilibrium states are not at a new
Buchdahl radius. For the second (Case B) then, we also keep $\beta=-1/3$, but now set
$\alpha$ via the mass dependent relationship (\ref{alpha_fix}) that the linear
analysis identified as being necessary to keep the asymptotic bubble's radius at its Buchdahl value.
We adopt the simpler
viscous equations (\ref{RHOGdot}) with $\tau_e=\tau_p=0$, and $\zeta=0.1$ and, when employing the alternative
flux option, we adopt $\tau_u = 2\times 10^{-6} m$.  These values of $\zeta, \tau_u$ are not special; the former are sufficiently
small to play only a secondary role in the dynamics; the latter imply a short time for the gas temperature
to approach its corresponding Unruh value and can be chosen up to $100$ times larger and still give essentially the same qualitative behavior\footnote{
Even larger values produce a solution which is quite sensitive to this choice; lower ones give the same behavior but if significantly
smaller leads to a stiff equation requiring a more delicate numerical treatment.}

Before illustrating the bubble's behavior when perturbed, we note that there is a maximum amplitude 
of the perturbing pulse (for reasonable choices of parameters $\{\tau^a_{\xi},\sigma_{\xi}\}$), that if exceeded
(some of) the equations become singular. This singular behavior takes place when $\xi_S \approx m^{-1}$,
which induces $a_R \rightarrow 0$ and $\dot m \rightarrow 1$, suggesting the bubble's growth
approaches the speed of light and the effectively classical description of the bubble's internal dynamics ceases to
make sense. As reference, for a perturbation with $\sigma_{\xi} \simeq m$, the largest mass change one can achieve
is of $\approx 12\%$ after the two interactions.
In what follows, we restrict to slightly lower values to avoid this situation. We consider a
bubble with initial mass $m=5000$ and choose the amplitude of the perturbation such that, after two 
perturbing episodes, the net relative change of the mass is $\Delta m/m = 0.2 \times 10^{-n}$
with $n=3..5$ and. To more clearly illustrate the asymptotic state of the solution, and its agreement
(or lack thereof) with a Buchdahl state, we normalize each plotted quantity either by the (instantaneous) value
expected for a Buchdahl solution, or by the initial value of that quantity. Further, we also
normalize by the inverse of the relative change in mass to more clearly compare with different chosen amplitudes.

\begin{figure}
\includegraphics[width=3.3in]{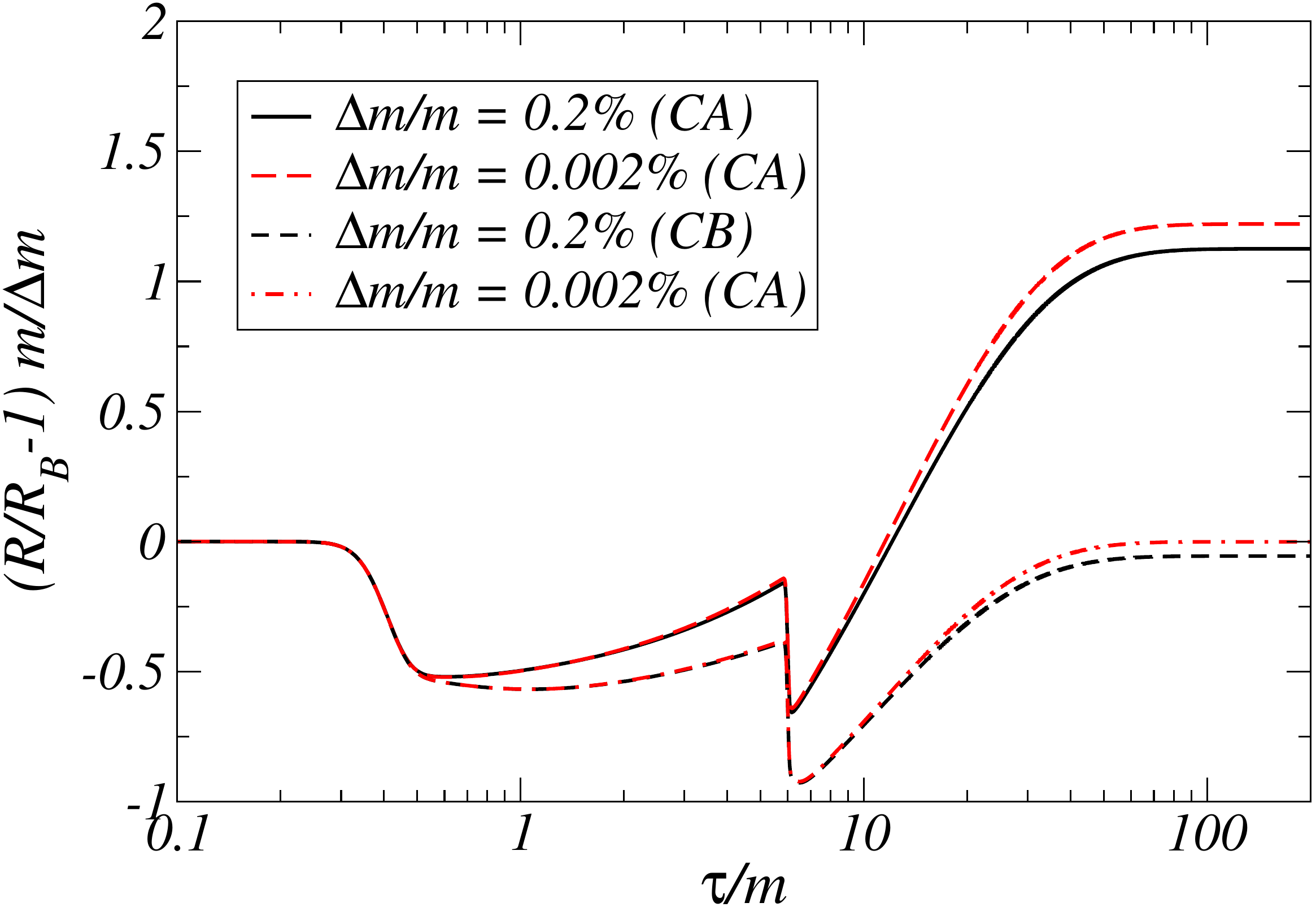}
\includegraphics[width=3.3in]{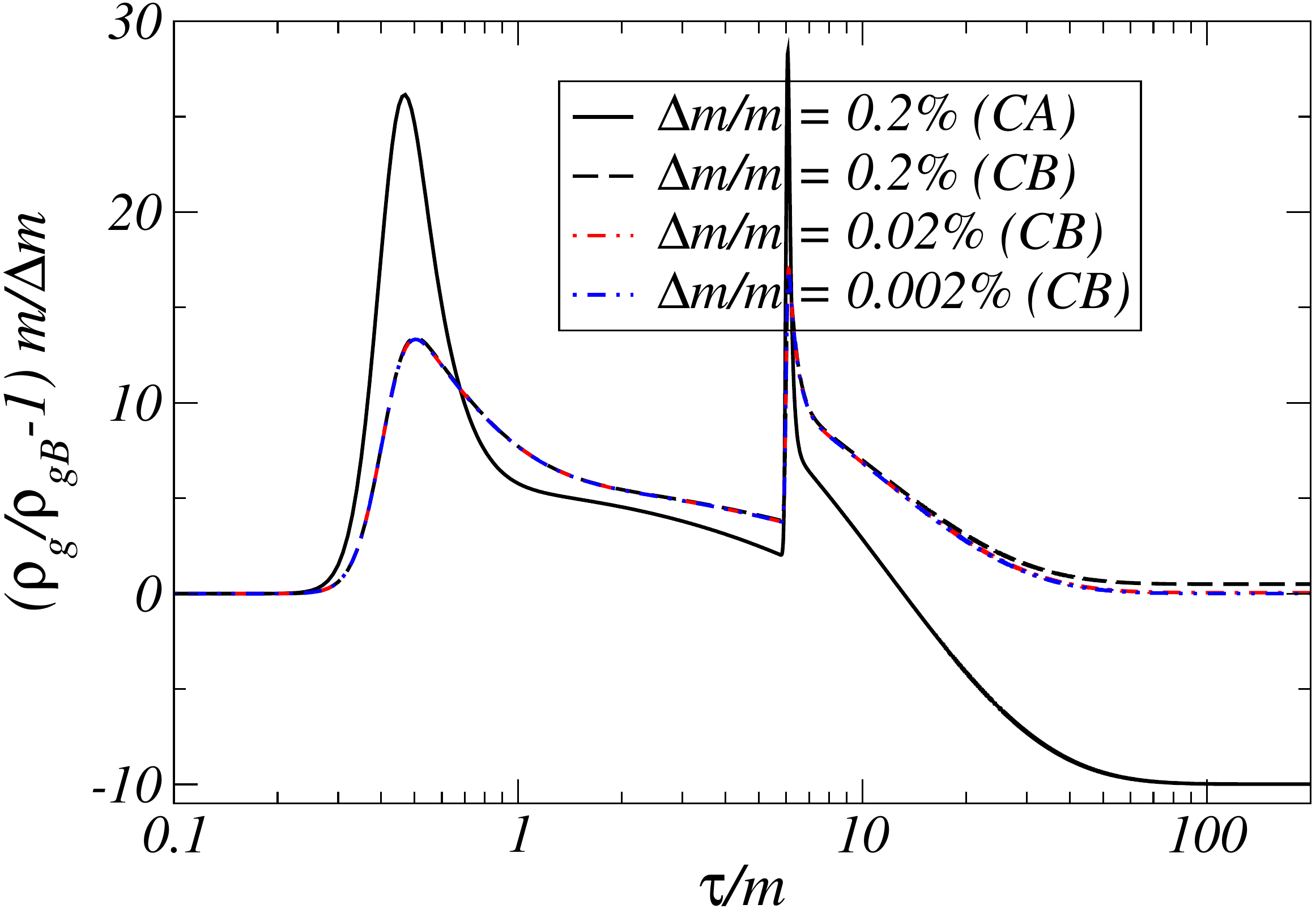}
\caption{
Normalized radius (top panel) and gas density (bottom panel) from evolutions
of a black bubble undergoing two distinct accretion episodes using the relaxation approach (\ref{j_u_def}) (with $m=5000$, $\ell=1$, 
$\tau_u=0.01$ and $\zeta=0.1$). In the top panel, four solutions are presented
corresponding to relative mass changes of $\Delta m/m=0.2, 0.002\%$ for each case. Case A asymptotes
to a non-Buchdahl yet stationary solution, while Case B converges to a Buchdahl state with a subleading
correction that goes to zero with $\Delta m$.
In the bottom panel, results corresponding to a mass change of $\Delta m/m=0.2 \%$ for case A, and $\Delta m/m=0.2, 0.02, 0.002\%$
for Case B are shown. Case A asymptotes to a stationary solution distinct from the Buchdahl one, while Case B converges to a Buchdahl state
in a similar manner with $\Delta m$ as the radius.
({\em Note that both accretion episodes are of the same duration; that the second looks
so abrupt is due to the logarithmic scale used for the time axis.})}
\label{fig:radius_highII}
\end{figure}

First, Fig.~\ref{fig:radius_highII} shows the behavior of radius and gas density vs ($\tau/m$). 
For both curves, we normalize them with respect to the corresponding quantities evaluated for
the equilibrium solution with mass corresponding to the bubble's instantaneous mass, and also
by the inverse of the relative mass change ($\Delta m/m)$. As can be appreciated from the figure,
while the late time solution for both cases is stationary, for Case A this does not correspond to
a Buchdahl state. On the
other hand, Case B shows both quantities converging to zero (the Buchdahl state) linearly with $\Delta m$.

\begin{figure}
\includegraphics[width=3.3in]{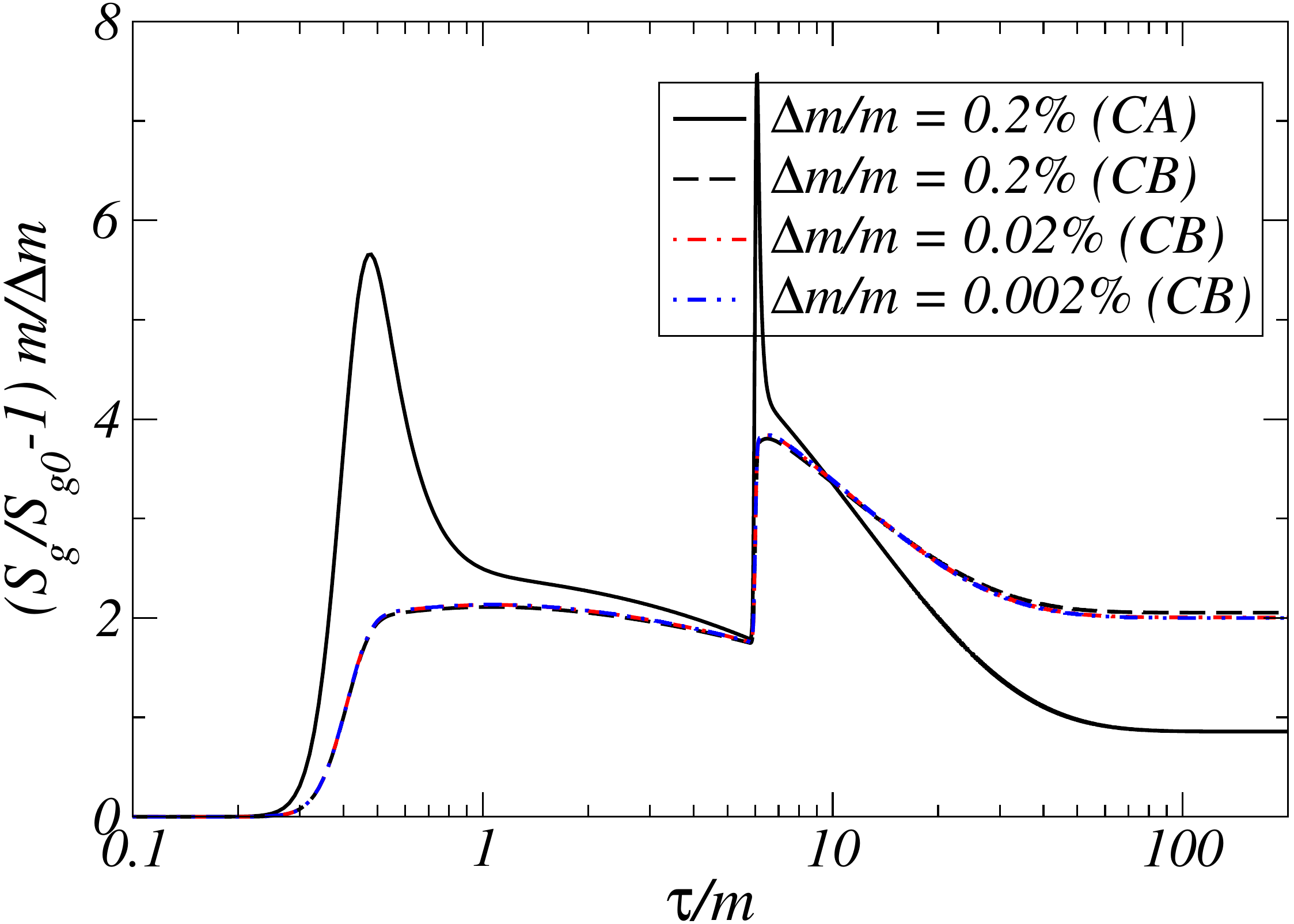}

\vspace{0.2in}
\includegraphics[width=3.3in]{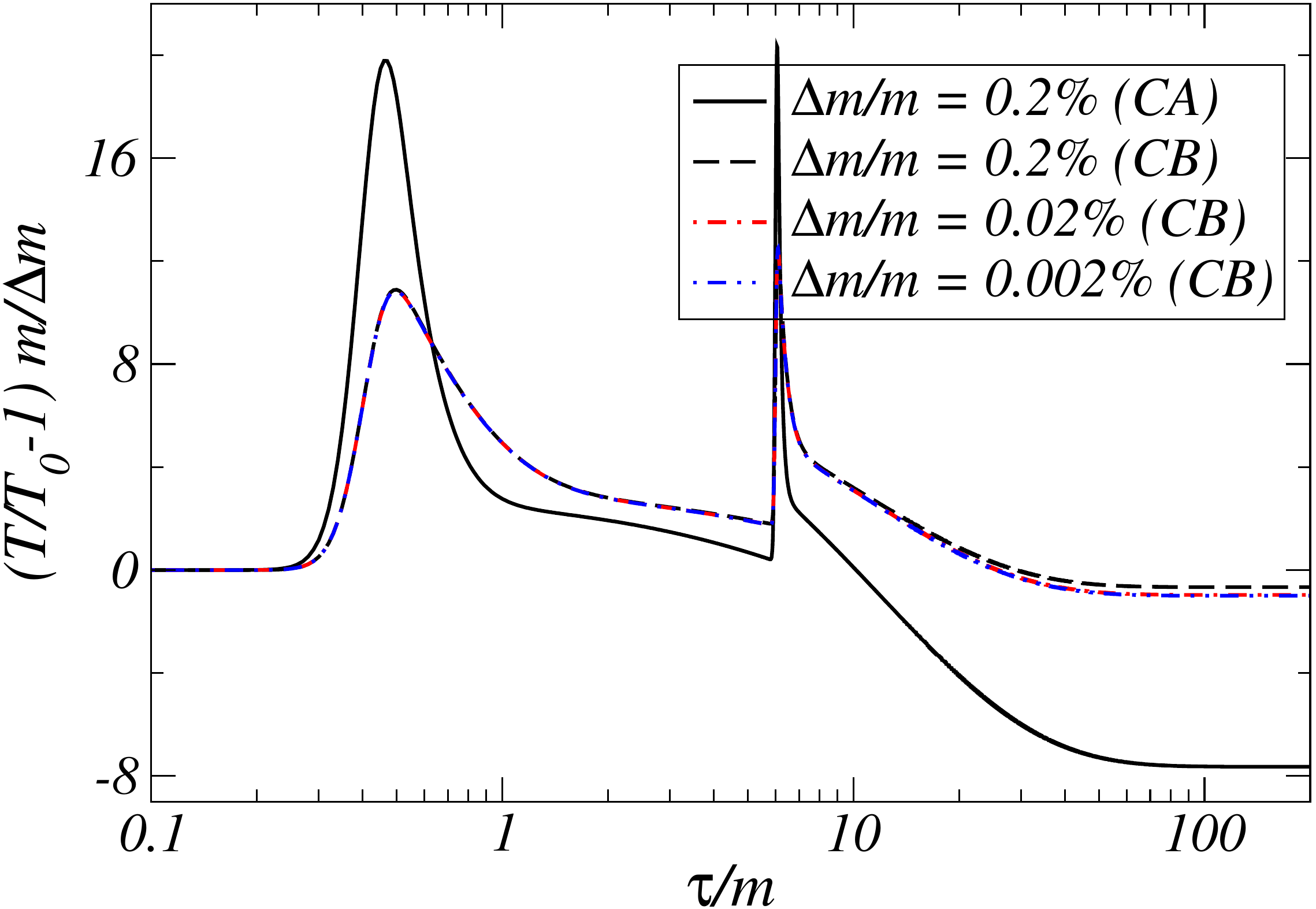}
\caption{
Entropy (top panel), and temperature (bottom panel), 
normalized with respect to their initial values, from evolutions
of a black bubble undergoing two distinct accretion episodes. Four runs
are shown using the relaxation approach (\ref{j_u_def}) (with $m=5000$, $\ell=1$, 
$\tau_u=0.01$ and $\zeta=0.1$) for Case A (with a relative change of mass $\Delta m/m=0.2\%$)
and Case B (with a relative change of mass $\Delta m/m=0.2\%, 0.02\%, 0.002\%$).
}
\label{fig:entropy_tauhII}
\end{figure}

Further insights into the dynamical behavior can be observed in Fig.\ref{fig:entropy_tauhII} which shows
the gas entropy  and the temperature (normalized by the initial temperature). The entropy shows a net increase from the
initial state to the final stationary solution, but as the interaction with
the perturbation takes place,  it shows a transient non-monotonic behavior. Comparing the net entropy
change (which can be consistently defined as the initial and final states are stationary)
indicates Case B has a larger final entropy than Case A. Quantitatively,
we find the net change of entropy from the initial state to the final equilibrium one 
is $\Delta S \approx C_{S_i} S_{g0} (\Delta m/m)$
with $C_{S_A}\approx 0.85, C_{S_B} \approx 2$. Recalling the gas entropy is $S_g =\rho_g R^2 T^{-1}$, and that for a state consistent with Buchdahl $\rho_g \propto R^{-1}$
for large masses, the value obtained for $C_{S_B}$ is the expected one for a Buchdahl state.
We note in passing, that one can choose values for
$\{ \alpha, \beta \}$ that guarantee a monotonic growth of gas entropy, but unreasonably large values of the dissipation parameter $\zeta$ would be required for stability. 
Finally, the temperature indeed shows the expected reduction in value as the bubble grows, exhibiting
a transient behavior as the interactions take place. Its asymptotic value denotes a change that can be approximated
by $\Delta T \approx C_{T_i}  T_0  (\Delta m/m)$, with $C_{T_A}\approx -7.5, C_{T_B} \approx -1$; the latter value corresponds to the
expected one for a Buchdahl final state.\\

\subsection{Numerical results from the Einstein-Klein-Gordon-Hydrodynamic system}\label{diss_EKG_results}

Since the ODE model can quickly and accurately study the stability of large 
black bubbles, we have, as demonstrated with some examples in the previous section, used that 
to map out black bubble matter properties that lead to stable configurations in spherical
symmetry. Here then, in the next two subsections we show a couple of results from the EKGH system
to illustrate some issues that would need to be addressed in future
studies exploring black bubbles beyond spherical symmetry. In the last subsection
we explore evolution of scalar fields on a fixed black bubble background, which
is possible with the EKGH code for long time scales and up to modest values
of the internal cosmological scale $\ell$.

\subsubsection{Accuracy and convergence}

One of the
issues limiting the EKGH code is related to accuracy : in this first attempt to model singular
layers in a PDE code we have sacrificed higher order convergence for the sake
of simplicity. That would not have been much of an issue if the stability 
of black bubbles did not depend so sensitively on the scales in the problem.
With a 1+1D code on a modern, single CPU machine we can 
evolve grids of up to $10^5$ points for a similar number of time steps
in about an hour of wall time. For small black bubbles, i.e. $\mb\leq1$,
even with a code that is only first order convergent, we can achieve
good accuracy over many shell light-crossing times. However, 
for reasons not entirely clear, though likely related to the ``mass amplification''
issue discussed in the following subsection, for $\mb\geq1$ 
the truncation error at a given resolution rapidly
increases with $\mb$, so much so that by $\mb\sim 10$
we cannot evolve for more than
of order a light-crossing time at the highest resolutions before $O(1)$
errors are reached (in mass conservation for example). Moreover,
with certain flux parameters there is a numerical
instability that seems to set in for large $\mb$ (or at least the
growth rate depends on $\mb$, and if present for smaller values is sufficiently
mild that we have not noticed any lack of convergence then).

\begin{figure}
\includegraphics[trim=0.3in 0.2in 0in 0in, width=3.8in]{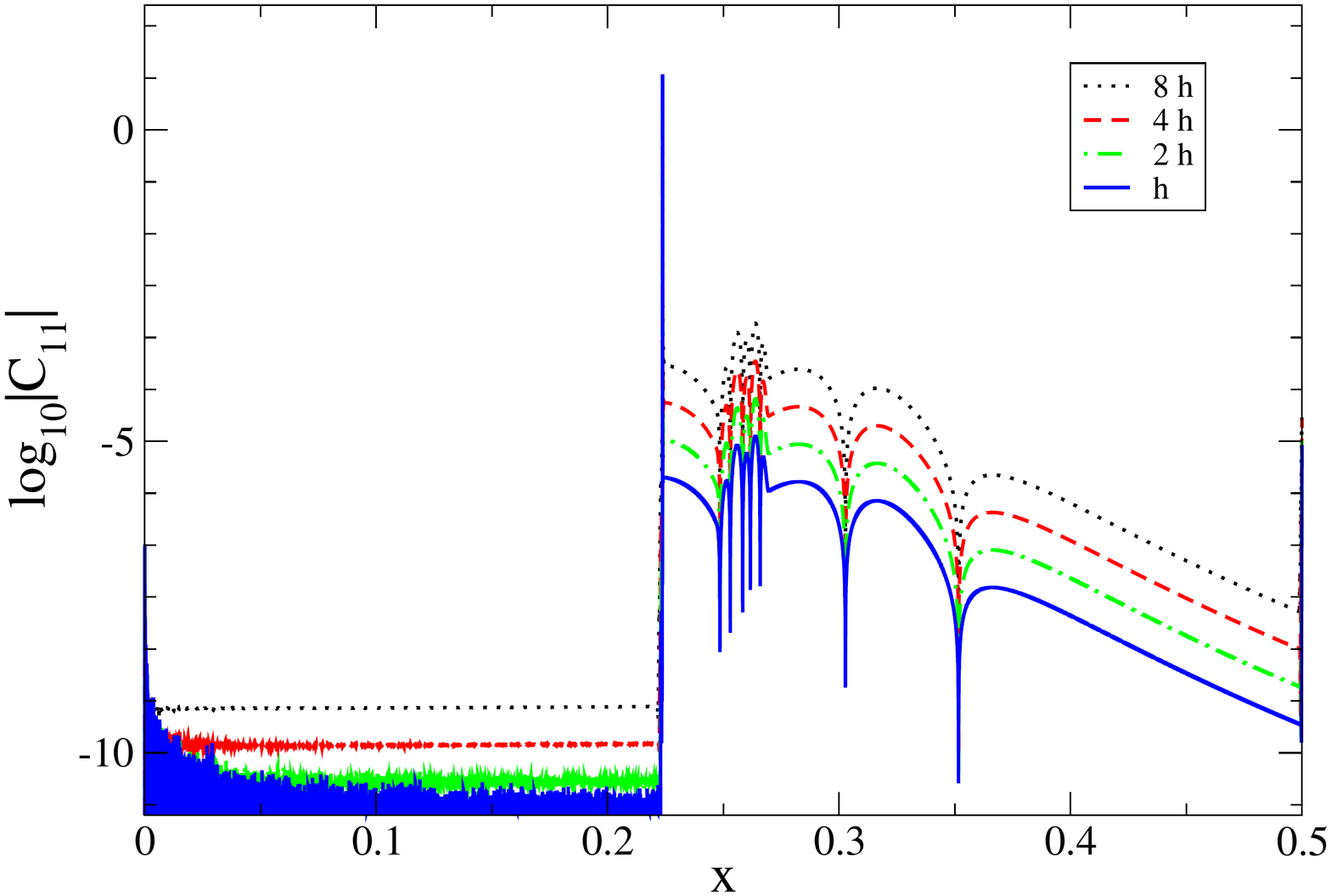}
\includegraphics[trim=0.3in 0.2in 0in 0in, width=3.8in]{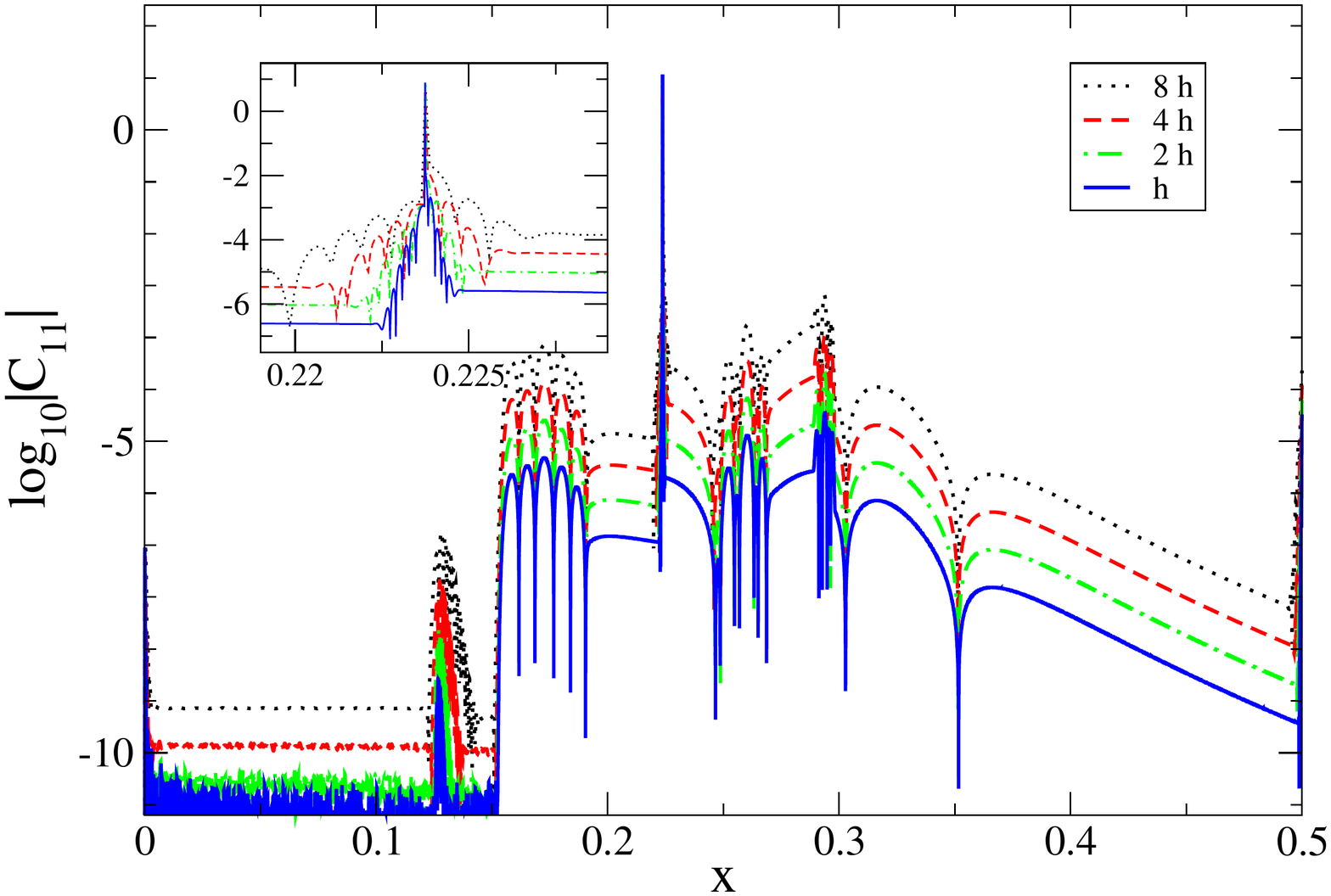}
\caption{
The residual of (\ref{C_const}) at $\tau/m=0.01$ (top panel) and $\tau/m=1.0$ (bottom panel),
for an $\mb=0.1$ black bubble perturbed with a non-interacting scalar field $\psi$
(using $\ell=1,\alpha=0.35,\beta=0,\tau_u=0.1, \zeta_0=1.0, \tau_e=\tau_p=0$),
where $\tau$ is proper time measured at the shell location. The finest resolution
mesh spacing is $\Delta x=h=0.5/32768$.
The shell is at $x\sim0.23$, corresponding (initially) to a proper radius $\rb=0.225 m$, while
the outer boundary $x=0.5$ corresponds to a proper radius $126m$.
At $\tau=0$ the scalar field pulse is centered at $x=0.27$, has a coordinate width of $0.04$ (\ref{sf_id})
and an amplitude so that it adds $\sim 0.002m$ to the mass of the spacetime (\ref{mass_aspect}).
The initial data is time symmetric, so half falls into the bubble (corresponds to the second set of
peaks out from the origin on the bottom panel---the smaller first peak is a transient
emanating from the shell location at $t=0$). The ``noise'' in the interior seems to be associated with the calculation reaching double-precision round-off error there.
}
\label{fig:C11_R0p225}
\end{figure}

\begin{figure}
\includegraphics[trim=0.3in 0.2in 0in 0in, width=3.8in]{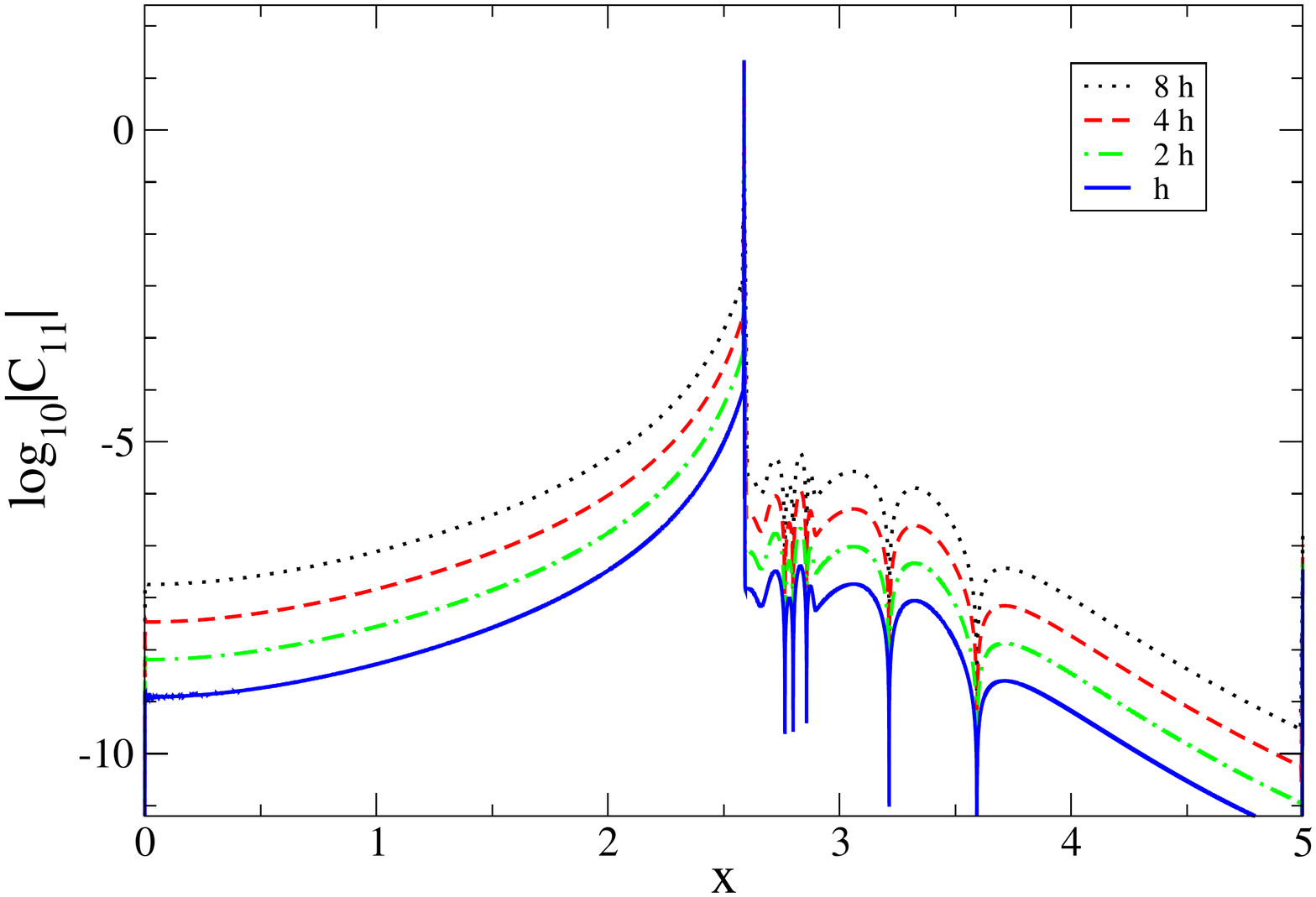}
\includegraphics[trim=0.3in 0.2in 0in 0in, width=3.8in]{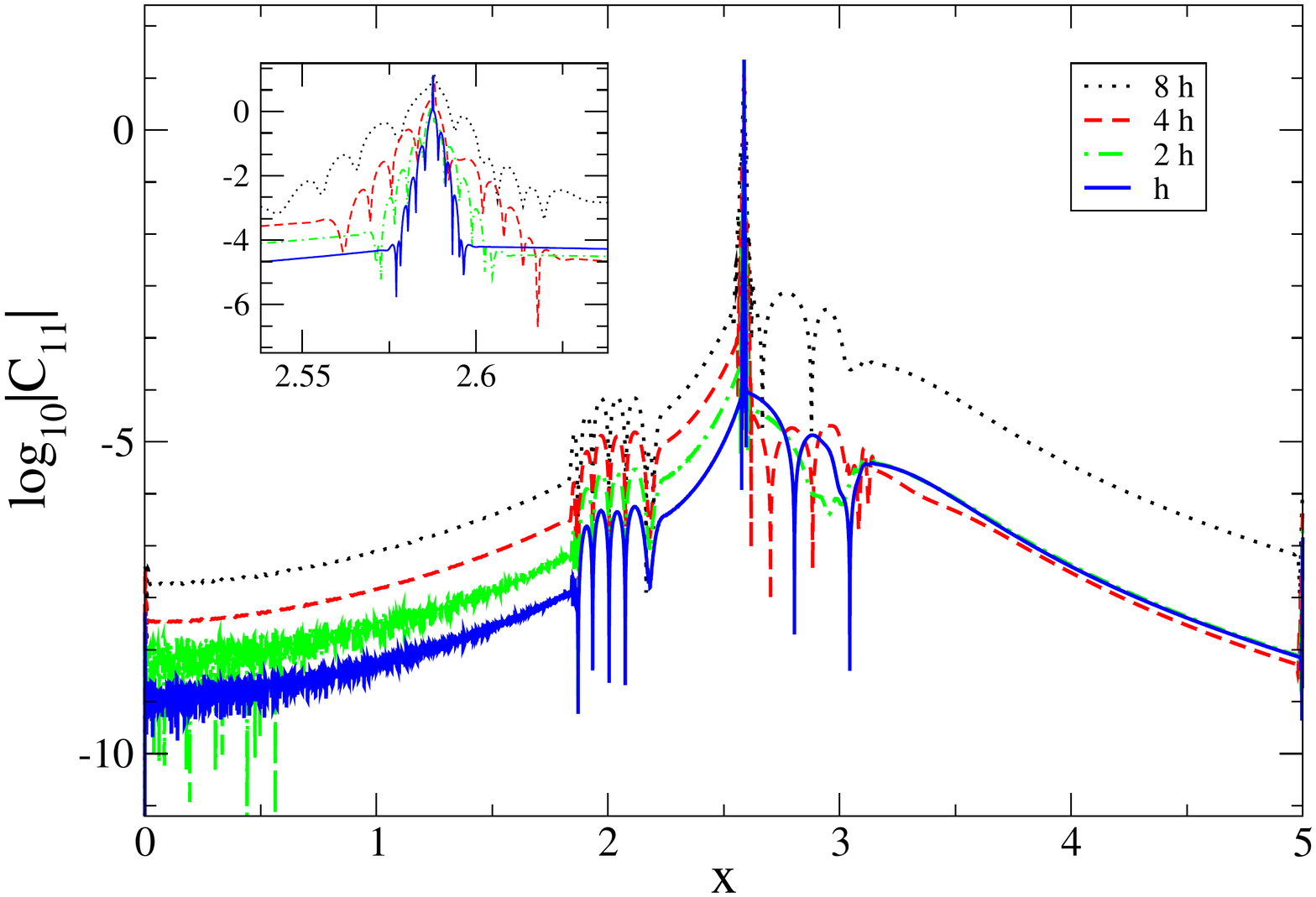}
\caption{
The residual of (\ref{C_const}) at $\tau/m=0.01$ (top panel) and $\tau/m=1.0$ (bottom panel),
similar to the case shown in Fig.\ref{fig:C11_R0p225}, but here 
for an $\mb=10$ black bubble, and the finest resolution mesh spacing is $\Delta x=h=5.0/32768$.
The shell is at $x\sim2.5$, corresponding (initially) to a proper radius $\rb=22.5 m$, while
the outer boundary $x=5.0$ corresponds to a proper radius $161m$.
At $\tau=0$ the scalar field pulse is centered at $x=2.8$, has a coordinate width of $0.4$ (\ref{sf_id})
and an amplitude so that it adds $\sim 0.002m$ to the mass of the spacetime (\ref{mass_aspect}).
In comparison to Fig.\ref{fig:C11_R0p225}, notice the different magnitudes
of the residuals. In particular in this case there is rapid growth of the residual
exterior to the bubble with time, and moreover it oscillates on a timescale
of order $\tau/m$---that the three higher resolutions seem to be the same
at large radii is mostly coincidence as the oscillations happen to overlap at $\tau/m=1.0$
(though there is also some deterioration of the rate of convergence, which does
happen on such short time scales for these large mass cases).
}
\label{fig:C11_R22p5}
\end{figure}

Figs. \ref{fig:C11_R0p225} and \ref{fig:C11_R22p5} show examples of convergence for two 
different mass black bubbles, $m\ell=0.1$ and $m\ell=10$ respectively, perturbed
with a non-interacting scalar field $\psi$
(these are also the two outlier
cases shown in Fig.\ref{fig:dmi_dme_vs_m0_mb2p24} below).
In both cases, after the ingoing component of the scalar field propagates
across the shell, this perturbation results in a change of the mass aspect $m(r,t)$,
defined via the following generalization of the Misner-Sharp mass\cite{Misner:1964je}
\be\label{mass_aspect}
1-2 m/\rb + \Theta(R-r)\ell^2 \bar{r}^2/3 \equiv \nabla^b \bar{r} \nabla_b \bar{r},
\ee
of $\sim0.1\%$ evaluated just exterior to the bubble location $R(t)$ 
(the net initial energy of the scalar field is roughly
twice this, with the other half propagating outward).
In the above $\bar{r}(r,t)\equiv \sqrt{{\bf A}(r,t)/4\pi}$ is areal radius.

What is shown in Figs. \ref{fig:C11_R0p225}-\ref{fig:C11_R22p5} are residuals of the 
constraint $C_{11}$ (\ref{C_const}) 
(i.e. the left hand minus right hand
side of it) evaluated pointwise across the grid using centered, second order accurate finite 
difference stencils, at two times during the evolution. With our mapping of the
shell to a constant location in $x=x_0$, we have also fixed that location to be at a vertex
of the grid. Therefore, a consistent representation of the delta function appearing in (\ref{C_const}) 
is to use the piecewise linear function that goes from $0$ at $x_0-\Delta x$ to $1/(2\Delta x)$ at
$x_0$, and back to zero at $x+\Delta x$. Then, having evolved with a first order accurate finite volume
integration about $x_0$ (see Appendix~\ref{sec_wfs}), one only expects a consistent, convergent
scheme to show convergence of a residual to zero in an integrated
sense; specifically, $C_{11}(x_0,t)$ will evaluate to a finite function of time irrespective
of resolution, though adjacent points around it should converge to zero first order
in $\Delta x$. This can be seen in Figs.\ref{fig:C11_R0p225}-\ref{fig:C11_R22p5},
though we do initially have second order convergence away from $x_0$, and interestingly
the region around $x=x_0$ that drops to first order with time grows quite slowly
compared to the characteristic speed of the system.

The main point we want to illustrate with those figures is how much
larger the truncation error is for large ($\mb=10$ in Fig.\ref{fig:C11_R22p5}) 
vs small ($\mb=0.1$ in Fig.\ref{fig:C11_R0p225}) mass bubbles. Note in particular
the interior region, which is empty AdS to begin with, while in the exterior
region by $\tau/m \sim 1$ the truncation error has grown to be of
comparable magnitude for the $\mb=10$ case, and we are beginning to loose convergence there.

\subsubsection{Interior energy}
The second problem affecting the EKGH system evolutions is related to a physical issue, in 
that in the $m\ell\gg1$
limit the bubble is very ``close'' to what would be the AdS boundary from the interior
spaces' perspective. One consequence of this is when we perturb the shell with a small,
exterior non-interacting scalar field pulse, as it crosses the shell it is very strongly
``blue shifted''. So in terms of a geometric mass (\ref{mass_aspect}) one can end up with
a lot inside the shell. In fact, it is even possible to perturb the shell
so that the interior mass ends up being larger than the asymptotic mass, 
and the shell acquires a negative gravitational mass. Such (and more modest cases)
typically form black holes in the interior; considering quantum effects
presumably such states will eventually tunnel to a larger, encompassing black bubble.

\begin{figure}
\includegraphics[width=3.5in]{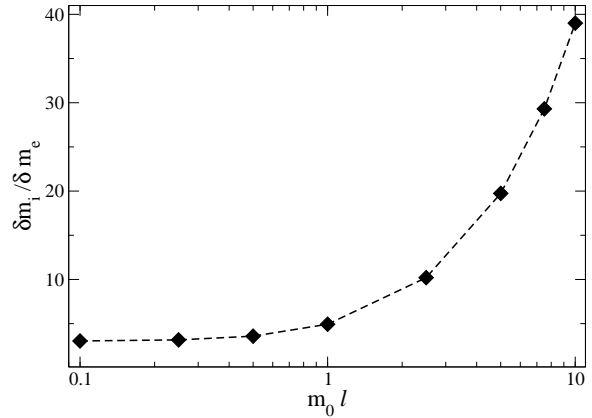}
\caption{
The change in interior mass $\delta m_i$ relative to the change measured
exterior to the shell $\delta m_e$ as a function of initial shell mass $m_0\ell$
(using $\ell=1,\alpha=0.35,\beta=0,\tau_u=0.1, \zeta_0=1.0,\tau_e=\tau_p=0$). From convergence
studies estimated uncertainties in $\delta m_i/\delta m_e$ are less than $1\%$ for all
points (the dashed line between the points is simply to guide the eye).
For all cases the parameters of the perturbing scalar field
were adjusted to give $\delta m_e \sim 0.001 m_0$. For larger masses $\delta m_i/\delta m_e$
grows linearly as a function of $m_0\ell$ (note that the figure has a logarithmic scale
for the x-axis).
}
\label{fig:dmi_dme_vs_m0_mb2p24}
\end{figure}

To illustrate this interior geometric-mass amplification, in Fig.\ref{fig:dmi_dme_vs_m0_mb2p24}
we plot the change in interior mass (\ref{mass_aspect}) $\delta m_i$, measured just inside 
the shell, as a fraction of the change in exterior mass $\delta m_e$, measured just outside the shell.
The $m_0 \ell=0.1,10$ cases are from the same evolutions shown above with the convergence
tests; the intermediate points are from similar runs with the perturbing
scalar field parameters adjusted to also give $\delta m_e \sim 0.001 m_0$ on a similar local time scale.
Note that the linear analysis shows that for these parameters
($\alpha=0.35,\beta=0,\tau_u=0.1,\zeta_0=1.0$) black bubbles with $m_0\ell\lesssim0.5$
are unstable, and this is confirmed by the code, though
for such relatively short interactions $\delta m_i/\delta m_e$ does not depend on 
the flux parameters (we have not found a single set of parameters that give
stable bubbles for both small and large masses). Also, since no energy is
directly exchanged with the shell matter, on these short time scales $\rho_g,\rho_\tau$ and
$\rho_s$ are roughly constant. The trend from the figure
on the large mass side is that  $\delta m_i/\delta m_e \approx m_0$ (e.g., for a similar
$0.1\%$ perturbation, cases with $m_0 \ell \gtrsim 1000$ will give negative gravitational
mass bubbles).

\subsubsection{Gradual release of internal energy}

For a rough estimate of the effect of internal energy, assuming it is not sufficient
to collapse to a black hole, nor trigger a quantum transition to a new black bubble
configuration, here we evolve a free, non-interacting scalar field $\psi$
on a black bubble background. With the PDE code we can run such cases
for many dynamical times, and up to modest values of $\ell$ of O(10). The specific
examples we show here choose an initial scalar field pulse of the form (\ref{sf_id}),
though use proper radius $\rb$ to define it to make for more meaningful
comparisons varying $\ell$ (the relationship between $r$ and $\rb$ in the
light-metric (\ref{metric}) coordinates depends strongly on $\ell$); we set
$R_\psi=3.5m$, $\Delta_\psi=m$ (and $m=1$ in all cases).

\begin{figure}
\includegraphics[trim=0.3in 0.2in 0in 0in, width=3.8in]{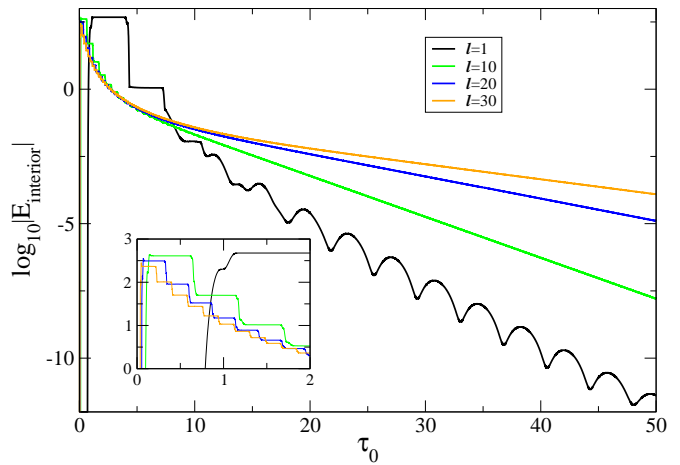}
\caption{
Logarithm of the integrated energy (\ref{SFE}) of the
non-interacting scalar field $\psi$ interior to the bubble, as a
function of central ($\rb=0$) proper time $\tau_0$. 
These are all from runs {\em without} back-reaction; i.e. the scalar
field is simply propagating on the black bubble background. Runs using
four different
values for the cosmological constant scale $\ell$ are shown, each with $m=1$, and an initial perturbation
of characteristic width $\Delta \rb=1$ centered outside the bubble at
a radius $3.5 m$.   The rate at which energy escapes clearly decreases with increasing $\ell$.
}
\label{fig:SFE}
\end{figure}

\begin{figure}
\includegraphics[trim=0.3in 0.2in 0in 0in, width=3.8in]{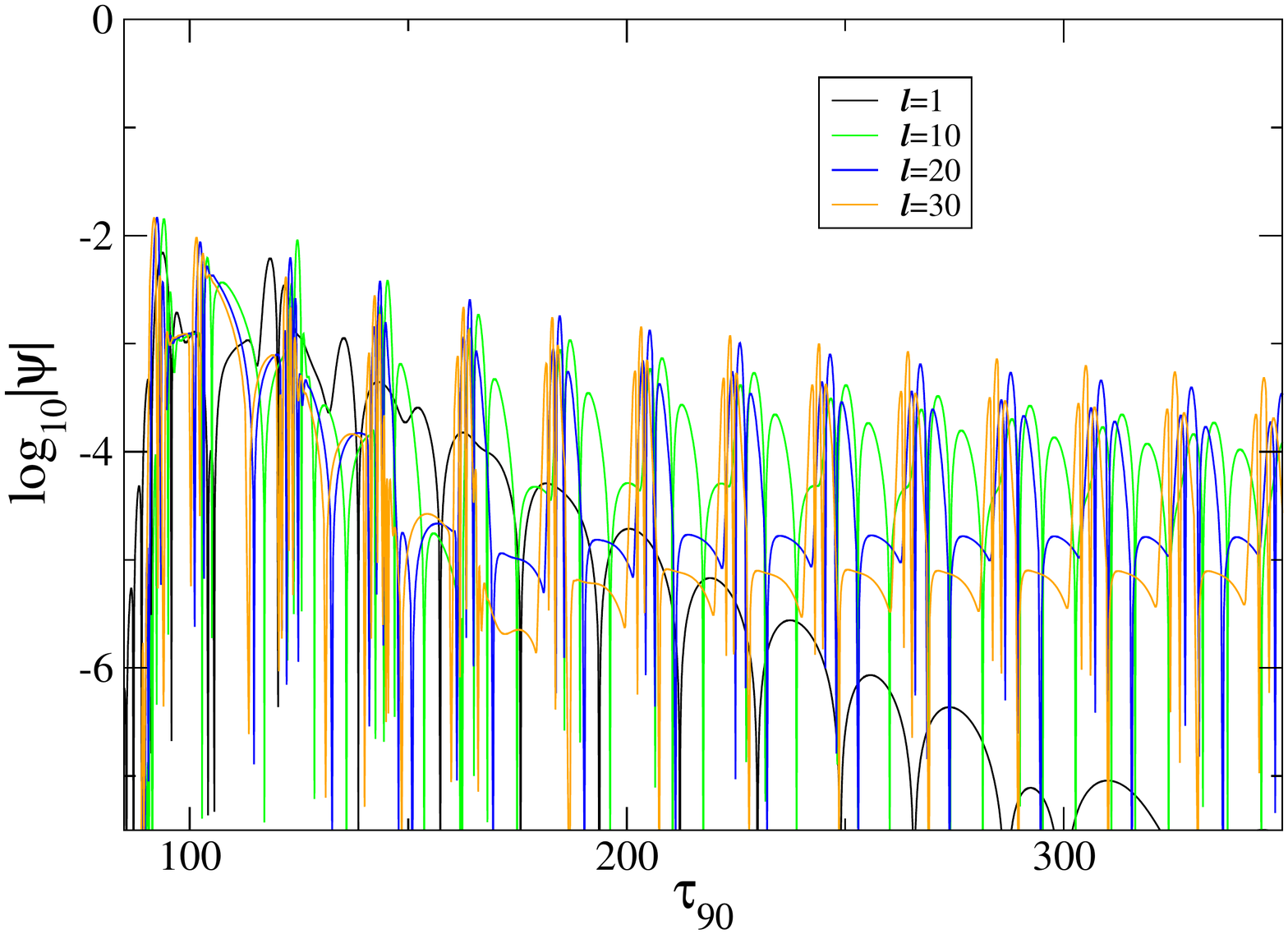}
\includegraphics[trim=0.3in 0.2in 0in 0in, width=3.8in]{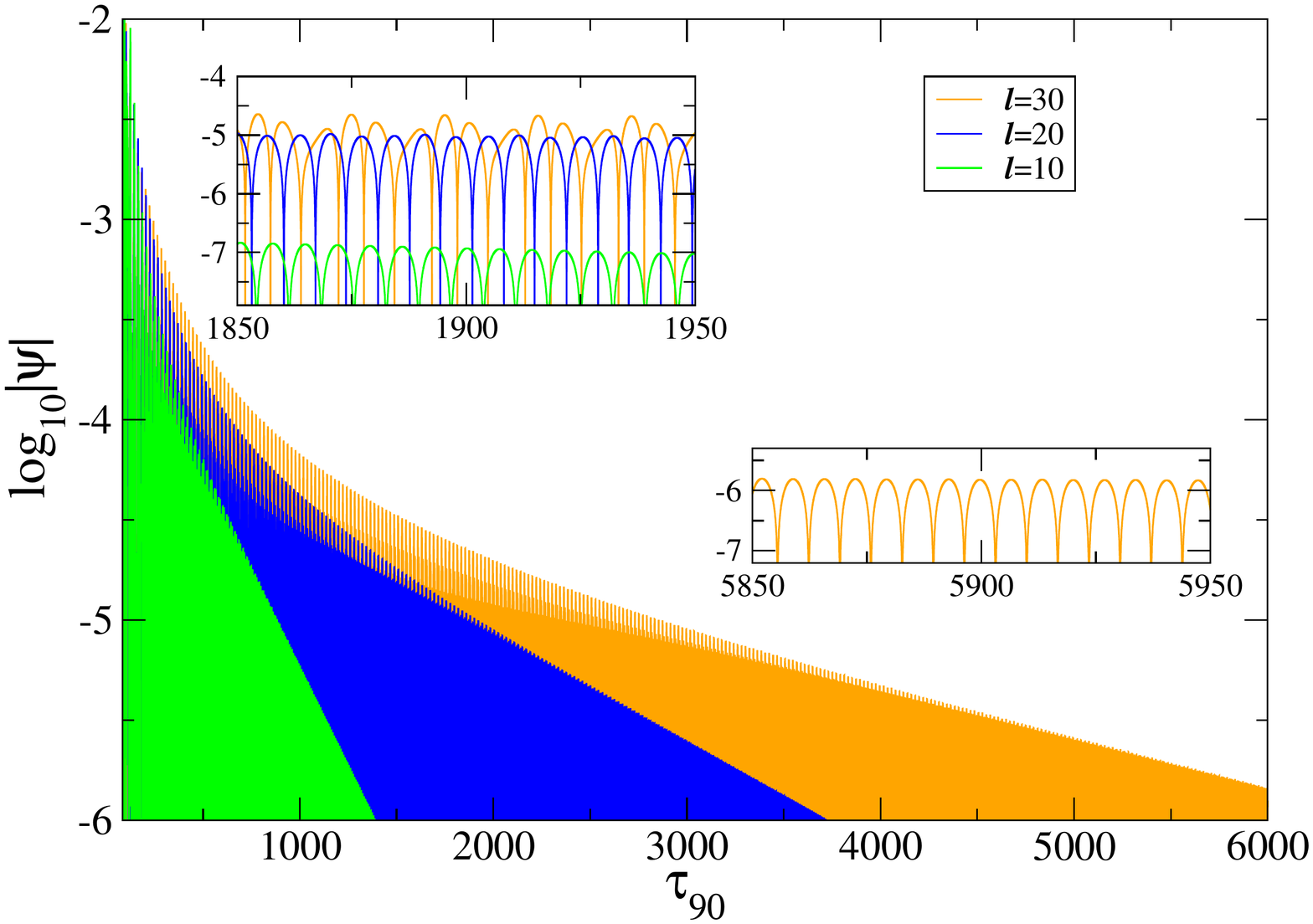}
\caption{
The amplitude of the scalar field measured at $\rb=90m$ ($m=1$) as a function
of proper time $\tau_{90}$ for a static observer at this location, for the same
cases shown in Fig.~\ref{fig:SFE}.
}
\label{fig:qnm12}
\end{figure}

The primary results are summarized in Figs.\ref{fig:SFE} and \ref{fig:qnm12}.
First, as shown in Fig.\ref{fig:SFE} the scalar field that crosses
into the bubble is partially trapped there, the more effectively
the larger $\ell$. Specifically, what is plotted there is the
integrated energy density interior to the bubble
\be\label{SFE}
E_{interior}(\tau_0) \equiv \int_{\rb=0}^{\rb=9 m/4} T_{ab} X^a n^b \sqrt{h} d^3 x,
\ee 
as a function of central proper time $\tau_0$, 
where $X^a = (\partial/\partial t)^a$ is the time-like Killing vector
of the static background, $n^a$ is the unit vector normal to $t={\rm const.}$ 
hypersurfaces, and $h$ is the determinant of the corresponding spatial
metric. On the background a similar quantity would be conserved if
the integral where carried out from $\rb=0$ to $\rb=\infty$.
The ``blocky'' nature of the curves at early times
is associated with the light-crossing time of the pulse interior to the
bubble, which decreases like $1/\ell$ with respect to the proper time
at the origin of AdS. Initially the pulse can be considered
to be a superposition of many AdS scalar field normal modes; the 
higher harmonics leak out more quickly, gradually leaving behind
the lower harmonics and a smoother late-time decay.

The reduction of energy within the AdS region can be understood
straightforwardly following the analysis of, e.g.~\cite{Calabrese_2003}, and
in Appendix~\ref{app_loss_rate} we outline such a calculation.
This shows that at late times when the fundamental mode dominates, and for large $\ell$,
one expects the interior energy to leak out via $\log E \sim \frac{-2\pi}{m^2\ell} \tau_0$; this scaling
with $\ell$ is consistent with the late time slopes
of the $\ell\ge 10$ curves shown in Fig. \ref{fig:SFE}.

In Fig.\ref{fig:qnm12} we show the imprint of this on the measured scalar radiation some distance
outside the bubble. 
A few interesting features
are apparent. Note the redshift between the oscillations with respect
to central proper time depicted in Fig. \ref{fig:SFE} and
the (near) asymptotic proper time in Fig.\ref{fig:qnm12} (the same run time of $\tau_0=54m$ 
translates to $\tau_{90}\sim 2120m,4160m,6150m$ 
for the $\ell=10,20,30$ cases respectively, though the corresponding curves stop
below the lower y-axis limits of the figures). This means
the observed rate of energy loss scales like $1/\ell^2$, as opposed
to the $1/\ell$ measured with respect to interior central proper time (see Appendix~\ref{app_loss_rate} for more details). In terms of the externally observed frequency,
the redshift also almost exactly compensates for the increasing internal oscillation frequency
with $\ell$, and the frequency observed at late times in the exterior
is roughly independent of $\ell$ (see the insets on the bottom panel).
Specifically, the late time fundamental harmonic mode of a scalar field in AdS with frequency 
(relative to central proper time) $\omega_0\sim\sqrt{3}\ell$ is observed
at large radii redshifted to $\omega_\infty\sim 4/9/m$.

\begin{figure}
\includegraphics[trim=0.3in 0.2in 0in 0in, width=3.8in]{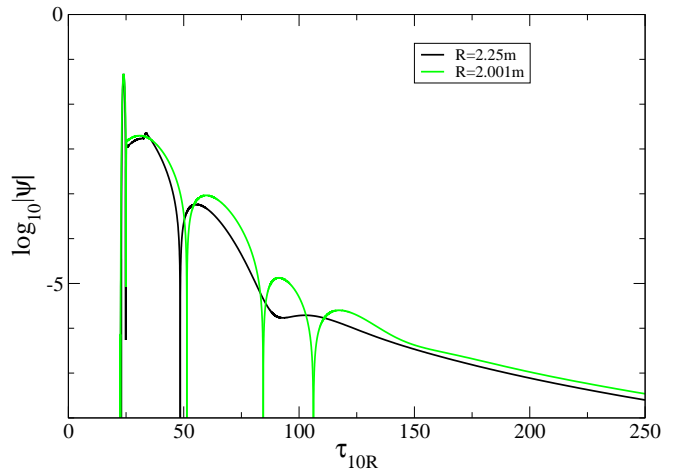}
\caption{
The amplitude of the scalar field measured at $\rb=10 R$ ($m=1$) as a function
of proper time there ($\tau_{10R}$), for similar initial data as
depicted for the runs in Fig.~\ref{fig:qnm12}, but here with
perfectly absorbing boundary conditions so that no scalar field enters
the bubble (hence $\ell$ is irrelevant). For the black curve the bubble
is at the canonical Buchdalh radius, while for the green it is
at $\rb=2.001m$, to mimic a black hole (and of course, such a bubble will
be unstable if back-reaction were included). Since the measurement
radii are at slightly different locations, one curve was shifted
in time to align the profiles at peak amplitude for ease of comparison.
This would likewise affect the relative amplitudes, which has not
been corrected for, though here we more want to emphasize the slight
shift in frequency and number of quasi-normal oscillations visible
before essentially the same power-law decay sets in.
}
\label{fig:qnm3}
\end{figure}

Finally in Fig.~\ref{fig:qnm3}, for comparison we show two similar non-backreacting runs,
but now using the accreting scalar with perfectly absorbing boundary conditions.
The first is the usual black bubble case at the Buchdahl radius, while for 
the second the radius has been set to $\rb=2.001$
to mimic a black hole (we cannot set the boundary at exactly $\rb=2$, as the
light-like coordinates become singular then). The results are qualitatively
similar, though do differ in detail, suggesting that the early time gravitational wave
signal from black bubble formation will be similar to the black hole case, 
yet distinguishable with a precise enough measurement. \\

We discuss some of the potential observational consequences of this in the next section.

\section{Discussion}\label{discussion}
In this work, we have taken first steps toward seriously considering the non-linear classical dynamics
of shell-like black hole mimickers (or ECO's --- Exotic Compact Objects). We formulated the problem 
within a fairly general framework that does not rely on symmetries of a single, isolated ECO,
though for simplicity in a concrete example we restricted to spherically symmetry. Similarly,
the novel techniques we introduced to implement this in a code were designed with
application beyond spherical symmetry in mind.

The particular model ECO we studied are the AdS black bubbles of ~\cite{Danielsson:2017riq}. 
This model is motivated by string theory, and the initial investigations in ~\cite{Danielsson:2017riq}
suggested they are stable---a crucial requirement for any astrophysically viable ECO.
An important physical ingredient for stable black bubbles is an internal
interaction between the matter components of the bubble that causally
reacts to external perturbations (such as accretion), keeping the bubble in an equilibrium
configuration. We found here that the original quasi-stationary flux prescription of ~\cite{Danielsson:2017riq} 
was inadequate to maintain stability in dynamical situations, and developed a two parameter
generalization of it. We identified regions of parameter space that do result
in stable black bubbles, at least for sufficiently slow accretion. Moreover,
within the space of stable bubbles we were able to find parameters that
guarantee (at the linear level for large black bubbles) that after
a dynamical episode the bubble relaxes to a new equilibrium black bubble, i.e.
it sits at the Buchdahl radius corresponding to its new mass.
Though we argued
that the new parameters can be considered ``natural'', we did not derive the new flux
prescription from fundamental considerations, which would be an avenue for future research.

For rapid accretion, namely
when a sizeable fraction of the mass of the bubble accretes within of order the lightcrossing
time, we do find that otherwise stable bubbles can collapse to black holes. However,
then the internal fluxes take on values that suggest the evolution is outside the realm
well described by the classical analysis. Likewise, anticipating what might
happen when two black bubbles merge (assuming our stability results carry to non-spherical
perturbations), a classical analysis should be valid during the inspiral up to a moment just before the 
actual merger. For the analogue black hole case, in terms of local physics a global apparent horizon 
suddenly forms that replaces the apparent horizons of the two separate black holes. Similarly, there 
could be a quantum transition from one to two bubbles occurring before the two bubbles actually touch. 
Classically, one could attempt to model this in the same way by replacing the two bubbles
with an encompassing single bubble. On the other hand, taking guidance from the way
event horizons fuse together, one may be able to engineer the interaction between two bubbles
so that at the instant of contact they similarly fuse into a single bubble.
In the extreme mass ratio limit where no trapped surfaces would form as the two bubbles get close and fuse,
the latter approach by itself could be an accurate approximation of the full quantum system
(i.e. it may be that tunneling only occurs with high probability if a trapped surface would have otherwise formed).

Based on our results of scalar fields propagating on black bubble backgrounds,
we can make some very speculative comments on observational consequences of black bubble
formation or mergers. First, regarding the gravitational wave analogue where the scalar
field is not absorbed by the bubble, this is unlikely to have observable consequences
if $1/\ell$ is close to the Planck length $L_p$, or a similarly small microscopic scale. For then,
as estimated in Appendix~\ref{app_loss_rate}, the internal energy is effectively trapped.
On the other  hand, one can take the perspective that we do not know what this scale
is, and one can use black hole merger data to constrain it, or detect an 
unexpectedly large scale. This would be similar to the recent analysis in~\cite{Westerweck:2021nue},
where they assumed there was an ECO with purely reflecting boundary conditions some distance
$\epsilon$ from the would be Schwarzschild radius, and the absence of a long-lived,
nearly monochromatic postmerger ringdown signal from GW150914 could constrain $\epsilon$ as
a function of the ringdown timescale. Note that they do not propose that their
ECO can actually reflect gravitational waves (which would require matter that is bizarre
even by the lax standards applied to ECOs), but that on long timescales the passage
through some interior geometry effectively looks like a reflection.
A black bubble with a large $1/\ell$ would
similarly produce a monochromatic latetime ringdown as illustrated in Sec.~\ref{diss_EKG_results},
however at a frequency related to the Buchdahl radius as opposed to the Schwarzschild radius
(and appropriately modified for rotation, as is necessary for GW150914 and was done in~\cite{Westerweck:2021nue}). 
This suggests that black bubbles could offer an interesting counter example
to the conclusions given in~\cite{Westerweck:2021nue}, namely that the absence of such
a signal can be used to infer that the geometry outside the remnant of GW150914
must be close to that of Kerr down to some microscopic distance $\epsilon$ close to the horizon.
I.e., for black bubbles,
absence of such a signal constrains the interior AdS
scale, but not macroscopic differences from Kerr in the exterior geometry.
To constrain the latter would require understanding the prompt emission at the time
of merger. 

The comments about a late-time post-merger signal in the previous paragraph assumed external
gravitational waves with wavelength of order the bubble radius propagate into the 
interior, and these essentially excite the lowest wavelength modes
of the $AdS$ interior. Such modes are very efficiently trapped there.
However, as suggested by equation (\ref{mode_jL}), if there are internal
quantum gravity processes that produce gravitational waves on small
scales $1/j \sim 1/\ell$, they would leak out on observationally interesting
timescales even for $1/\ell\sim L_p$. Also, if the mass-amplification effect
illustrated in Fig.\ref{fig:dmi_dme_vs_m0_mb2p24} would classically cause
a black hole to form in the interior, this will instead induce a tunneling event
(or interior energy may induce tunneling to a new bubble regardless of classical black hole formation),
and the arguments for the rate at which energy leaks out given in Appendix~\ref{app_loss_rate}
would be invalid.

On another observational front, to explore how EHT images of supermassive black
holes would change if they were supermassive black bubbles, it would
be interesting to understand magneto-hydrodynamic (MHD) accretion from realistic
models of accretion disks onto black bubbles. Back reaction is likely unimportant,
and though black bubble spin would be, a good indication of whether the EHT
could discriminate between black bubbles and black holes could be made using
an exterior Schwarzschild background to begin with, assuming sufficient control
of gastrophysical processes are at hand. For the black bubble/MHD interaction
a conservative approach would be to model it as perfectly absorbing, as with the
scalar field case studied here.

There are many directions for future numerical studies of black bubbles.
The most crucial would be to relax spherical symmetry to explore
stability to non-radial perturbations, and if stable, accretion of angular momentum
to uncover the rotating solutions. The fact that the bubble surface
is within the photon sphere of the spacetime suggests there may be long
timescale secular instabilities~\cite{Keir:2014oka,Cardoso:2014sna,Cunha:2017qtt}. 
Classically, this might be analogous
to the so-called weakly turbulent instability of AdS spacetime~\cite{Bizon:2011gg}, which 
certainly is also relevant for the black bubble interior. If so, the consequence
of the instability might ``merely'' be that trapped energy could eventually
form small black bubbles that merge with the larger one. For rotating
black bubbles, similar instabilities could be associated with the
presence of an exterior ergoregion~\cite{1978CMaPh..63..243F,Moschidis:2016zjy,Keir:2018hnv}. Also,
it would be interesting to investigate whether in such cases there could be
superradiant extraction
of rotational energy, which may lead to similar observational signatures
as the presence of ultra-light particles around rotating black holes (see e.g.~\cite{Brito:2015oca}).
Rotational energy may also be extracted if a Chandrasekhar-Friedmann-Schutz instability
operates in fluid shells~\cite{1970ApJ...161..561C,Friedman:1978hf} (it is
generic for rotating fluid stars in general relativity).

Regarding the physics of black bubbles, a next step would be
to investigate whether the ad-hoc flux model prescribed here can be justified
with more rigor. To fully capture the physics of the bubbles when they tunnel and merge will be challenging. 
It would require a significantly new conceptual understanding of tunneling in a time-dependent background, 
as well as the construction of methods capable to implement this numerically. 
This would also be pertinent to understanding how soon after a merger the current model
can be applied, which should adequately describe the late-time ringdown.

Last, we note that the AdS black bubbles we focused on are but one of many 
potential ECO models. In that context, we hope our study, both in terms
of the methods we have introduced and how we solved issues particular to black bubbles,
can serve as a guide to further develop related ECO models.  
Likewise, since the potential observable features indicated in this work 
can be traced back to key aspects of the model's fundamental building blocks, other
ECOs with similar structure should exhibit the same qualitative observational characteristics.
For instance, relating the late time quasi-monochromatic radiation frequency 
to a redshifted fundamental mode of the interior region, as well as connecting the amplitude of decay
to interior energy loss, should be broadly applicable to any shell-like ECO with a compact, leaky
interior.

\begin{acknowledgments}
We thank Will East, Anna Ijjas, Eric Poisson and Paul Steinhardt for discussions. 
FP, UD, LL thank the Simons Foundation for support to attend a workshop where this work was initiated.
We acknowledge support from NSF under grant PHY-1912171 (FP), the Simons Foundation (FP), 
NSERC (LL) and CIFAR (FP, LL).
Research at Perimeter Institute is supported by the Government of Canada and by the Province of Ontario
through the Ministry of Research, Innovation and Science.
\end{acknowledgments}

\appendix

\section{Dynamical effects on proper acceleration}\label{dynamicalproperacc}
The analysis of \cite{Danielsson:2017riq} found that a black bubble is stable
to what are effectively quasi-stationary perturbations. Specifically,
they considered the proper acceleration of an exterior, stationary observer,
\be\label{as_eqn}
a_{s} = \frac{f'}{2 \sqrt{f}}
\ee
with $f=1-2m/\rb$ in Schwarzschild coordinates, $\rb$ is areal radius,
and here we use $\ '\equiv d/d\rb$.
The fractional change in acceleration that goes into the expression
for the flux (\ref{flux_eqn}) was then defined to be
$\dot{a}_s/a_s = V a'_s/a_s$, where $V=\dot{R}(\tau)$ is the shell velocity,
and $\dot{\ }\equiv d/d\tau$, with $\tau$ proper time along the shell trajectory.

However, in a dynamical situation there are additional terms that
appear in the expression for the 4-acceleration $a_R$, and it turns out these
can counter the effect of the quasi-stationary term,
and in fact so much so that adding the corresponding flux $j$ can make the bubbles
more unstable than without a flux. 
To see this, we evaluate (\ref{four_a}) in Schwarzschild coordinates for a moving observer
with unit 4-velocity $u^a=dx^a(\tau)/d\tau$:
\be\label{a_gen}
a_R=\frac{1}{2} \frac{f'+2A}{\sqrt{f+V^2}},
\ee
with $A(\tau)=\dot{V}(\tau)$. This already hints at problems, as the 
acceleration can have an arbitrary sign irrespective of the motion of the shell, and
the ``wrong'' sign will hinder the ability of a flux $j$ based on (\ref{flux_eqn})
to return a perturbed shell to equilibrium. To see this more clearly,
and that the general flux expression (\ref{flux_eqn_G}) with appropriate parameters could mitigate this
problem, we compute (\ref{flux_eqn_G}) using (\ref{a_gen}). The full expression is lengthy and somewhat obscure;
to simplify we evaluate it to leading order in $V$ at the Buchdahl radius:
\bea
& &\frac{j m}{\rho_g}\Big|_{\Rb=9m/4} = \frac{243\alpha \bar{J} }{81 \bar{A} + 16}\nonumber\\
      & &-V\left[\frac{64(6\alpha-\beta) + 81 (4[8\alpha-\beta]+81 \alpha \bar{A})\bar{A} }{3(81 \bar{A} + 16)}\right]\nonumber\\
      & & + O(V^2),
      \label{j_br_1}
\eea
where we introduced the jerk $J(\tau)=\dot{A}(\tau)$, and rescaled the jerk and acceleration
to given dimensionless quantities via $\bar{J}\equiv m^2 J$ and $\bar{A}\equiv m A$.
The quasi-stationary case $j_s$ is this expression with $\bar{A}=\bar{J}=0$ and $\alpha=\beta=1$
\be
\frac{j_s m}{\rho_g}\Big|_{\rb=9m/4} = -\frac{20 V}{3}.
\ee
Thus, for initial data where $V$ is small, but $A$ and $J$ are zero, the analysis
in ~\cite{Danielsson:2017riq} should hold, and the flux (\ref{j_br_1}) should start to counter the motion
of the shell.
However, this is not a generic
perturbation, and perhaps a more ``realistic'' perturbation for a presumed stable shell
would be the opposite case,
i.e., we imagine a black bubble has formed and settled down to a stationary spacetime,
then we throw in an external perturbation. In that case the first term in (\ref{j_br_1})
will dominate the flux, and this does not generically have the correct sign.
Equation (\ref{j_br_1}) also suggests that a simple alternative stable prescription
for triggering the internal fluxes is one based entirely on local changes to the area ($\beta\neq 0$)
and not the Unruh temperature ($\alpha=0$),
if $\beta$ is sufficiently negative.

\section{Linear perturbation analysis}\label{vac_sd}
In lieu of a full stability analysis, we will check whether stability is at least possible 
by seeking periodic solutions of the linearized equations for the simplified
shell model presented in Sec.~\ref{diss_model}. We start with the following ansatz
\bea
R(\tau)&=&R_0 + \delta_R \cdot e^{i \omega \tau},\nonumber\\
\rho_i(\tau)&=&\rho_{i0}+\delta_i \cdot e^{i \omega \tau},\label{ansatz}
\eea
where $R_0=9m/4$, $\rho_{i0}$ are the equilibrium matter
parameters for $i\in (g,s,\tau)$, and $\delta_R,\delta_i$ are the magnitudes
of a small perturbation. Here we do not include any external fluxes,
but assume they were responsible for creating these perturbations.

Consider 4 options for the flux term $j$:
$j=j_0=0$; $j=j_s$ from (\ref{flux_eqn}) with $a_R$ given by the quasi-stationary case
(\ref{as_eqn}); $j=j_d$ from the dynamical flux (\ref{flux_eqn_G}) with $a_R$
given by the full expression (\ref{aR_full}); $j=j_u$ 
using the alternative prescription for the dynamical flux
(\ref{j_u_def}) that explicitly introduces the temperate
$T$ and a corresponding relaxation to the Unruh temperate via (\ref{T_eqn_def}).
For the latter, we also adopt a similar ansatz for the temperature perturbation
\be
T(\tau)=T_0 + \delta_T \cdot e^{i \omega \tau},\\
\ee
with $T_0=8/(27\pi m)$. For dissipation, we assume $\zeta$ is a constant $\zeta_0$.
Plugging the above ansatz and flux options into the equations of motion,
and expanding to linear order in ($\delta_R,\delta_i$), a solution
consists of constraints on the amplitudes
of the matter (and temperature) perturbations $\delta_i$ ($\delta T$) in terms of the radial perturbation
$\delta_R$, and a relation $\omega(m,\ell)$. These 
are more conveniently expressed in terms of dimensionless variables
$\bar{m}=m\ell, \bar{\omega}=\omega/\ell, \bar{\tau}_u=\tau_u\ell$. 
We obtain for $j=0$:
\bea
\delta_{g,j_0}&=&- \delta_R\  \frac{4\ell^2}{81\pi\bar{m}^3}
                   \bigg[(\bar{m}-4\sqrt{3}/9)\nonumber\\
                   & &\hspace{0.5in} +\frac{4\bar{m}}{\sqrt{16+27\bar{m}^2}}\bigg] ,\\
\delta_{s,j_0}&=&- \delta_R\  \frac{16\sqrt{3} \ell^2}{729 \pi \bar{m}^3},\\
\delta_{\tau,j_0}&=&0;
\eea
for $j=j_s$:
\bea
\delta_{g,j_s}&=&6\ \delta_{g,j_0},\\
\delta_{s,j_s}&=& \delta_{s,j_0},\\
\delta_{\tau,j_s}&=&-5\ \delta_{g,j_0},
\eea
for $j=j_d$:
\bea
\delta_{g,j_d}&=&\left[1+6\alpha-\beta+\frac{729\alpha\bar{\omega}^2\bar{m}^2}{64}\right]\ \delta_{g,j_0},\\
\delta_{s,j_d}&=& \delta_{s,j_0},\\
\delta_{\tau,j_d}&=&-\left[6\alpha-\beta+\frac{729\alpha\bar{\omega}^2\bar{m}^2}{64}\right]\ \delta_{g,j_0},
\eea
and for $j=j_u$:
\bea
\delta_{g,j_u}&=&\frac{\delta_{g,j_d} - i \bar{\tau}_u\bar{\omega} (\beta-1) \delta_{g,j_0}}{1+i\bar{\tau}_u\bar{\omega}},\\
\delta_{s,j_u}&=& \delta_{s,j_0},\\
\delta_{\tau,j_u}&=&\frac{\delta_{\tau,j_d} + i \bar{\tau}_u\bar{\omega} \beta \delta_{g,j_0}}{1+i\bar{\tau}_u\bar{\omega}},\\
\delta_T&=& -\delta_R \frac{(128 + 243 \bar{m}^2\bar{\omega}^2)\ell^2}{162\pi\bar{m}^2 (1+i\bar{\tau}_u\bar{\omega})}.
\eea

The expressions for $\bar{\omega}$ are lengthy and not too illuminating by themselves, so
for simplicity we only show the more relevant large $\bar{m}$ limit:
\bea
j=0    &:&  \bar{\omega} \approx   \frac{32\pi\zeta_0}{27\bar{m}}  \left( i \pm  i \sqrt{ 1+9/32/(\pi\zeta_0)^2}\right),\\
j=j_s  &:&  \bar{\omega} \approx   \frac{32\pi\zeta_0}{27\bar{m}}  \left( i \pm  \sqrt{-1+27/(8\pi\zeta_0)^2}\right),
\eea
\bea
&j&=j_d,j_u  :  \bar{\omega} \approx    \frac{128\pi\zeta_0}{27(4-9\alpha)\bar{m}}\cdot\nonumber\\
\big(&i&\pm\vspace{-0.2in}\sqrt{-1+9(4-9\alpha)(6\alpha - \beta-2)/(16\pi\zeta_0)^2}\big). \label{jdju}
\eea
For $j=j_u$ there are 3 solutions if $\alpha\ne0$; the first two are identical
in the large mass limit to that of $j_d$ (\ref{jdju}), with the third given by
\be\label{ju_omega}
j=j_u : \bar{\omega} \approx \frac{i(4-9\alpha)}{4\bar{\tau}_u} -
                     \frac{i\alpha}{\bar{m}}\left[\frac{1}{\sqrt{3}\bar{\tau}_u} + \frac{64\pi\zeta_0}{3(4-9\alpha)}\right] \ \ \ (\alpha\ne0).
\ee
The zero flux ($j=0$) and canonical $(\alpha=1,\beta=1)$ dynamical flux $j_{d,u}$ cases always have at least one growing mode,
while the quasi-stationary flux $j_s$ is always damped\footnote{One could use the original 
quasi-stationary flux $j_s$ and achieve linearly stable black bubbles, at least in spherical symmetry. However this is a non-local flux, i.e.
a fluid element on the bubble needing to respond to a perturbation cannot, using any local
measurements of matter or spacetime properties, ``know'' what $j_s$ should be. Moreover, it is
unclear how \ref{as_eqn} could be extended beyond spherical symmetry even were one eager to adopt
non-local physics.}.
Various parameters can be found for the dynamical fluxes $j_{d,u}$ to give damped systems.
The third solution existing for the relaxation-based dynamical flux $j_u$ is always
stable for $\alpha<4/9$, and $\bar{m}$ sufficiently large that the second term
in (\ref{ju_omega}) is subdominant. 

\subsubsection{Particular solution}
In the analysis above we did not include any external flux,
assuming it was active prior to (say) $\tau=0$ to set up the perturbation,
after which one expects the solution to be given by some superposition
of the above modes. In this regard, one thing missing from the above ansatz
(\ref{ansatz}) are the arbitrary small perturbations of the initial
conditions that depend on the details of the prior external flux interaction.
It is straightforward to show that including such general initial
conditions requires adding a particular solution that simply
shifts the final radius and temperature (for damped, stable cases) 
by constants dependent on these initial parameters, but otherwise
does not affect any of the linear modes.

Similarly, if the perturbation caused some matter to flow to the
interior, and we model this as a small change $\delta m_i$ to the
interior mass, i.e. letting $f_L\equiv1+R(\tau)^2\ell^2/3-2\delta m_i/R(\tau)$,
we can solve the linear equations if we add the following constant correction
to $R_0$ in (\ref{ansatz})
\be
R_0\rightarrow R_0 + \frac{8}{81 \bar{m}^2} \delta m_i + O(1/\bar{m}^4).\\
\ee
(A corresponding correction to $T_0$ scales like $O(1/\bar{m}^4)$). 
This is a tiny correction to $R_0$, however, reversing the perspective,
a perturbation that leaks energy into the interior resulting in a small 
change $\delta R$ to the position of the bubble leads to a comparatively huge interior
mass parameter $\propto \bar{m}^2 \delta R$. It is not clear
that we can combine this with the result shown in Fig.\ref{fig:dmi_dme_vs_m0_mb2p24} where the increase
in interior mass comes from a scalar field interaction, and $\delta m_i \propto \bar{m}\delta m_e$ :
for small perturbations the scalar field will eventually escape, and for larger perturbations
where a black hole forms to trap the scalar field, a linear
analysis might not be warranted. Nevertheless, combining them for the case
where an interior black hole does form, this suggests a change
in radius (again for stable, damped cases) $\delta R \propto \delta m_e/\bar{m}$. 
In other words, this kind of perturbation, regardless of the flux parameters,
will lead to a new (classical) equilibrium position that is not exactly at the new Buchdahl radius.

\subsubsection{Impulse response}
For stable black bubbles, to determine what (if any) internal matter fluxes are capable 
of maintaining the bubble at the Buchdahl radius after an accretion episode $\xi_U(\tau)=\xi_S(\tau)\equiv\xi(\tau)$,
we consider the response of a bubble to an impulsive accretion event $\xi(\tau)=A \delta(\tau)$, with $A$ 
a constant amplitude parameter. If flux parameters
can be chosen to maintain such a condition for the impulsive response, then it should
likewise be maintained at the linear level for arbitrary accretion profiles $\xi(\tau)$.
Mathematically we can only make sense of a delta function source using the alternative
flux model (\ref{T_eqn_def}-\ref{j_u_def}); for simplicity we also only consider the simplified
dissipation model (\ref{RHOGdot}).

The first step is to integrate equations 
(\ref{Rdot}-\ref{V_tau_eom_J2},\ref{RHOSdot}-\ref{aR_full},\ref{RHOGdot},\ref{T_eqn_def}-\ref{j_u_def}) about $\tau=0$, 
with $\xi(\tau)=A \delta(\tau)$, to obtain the change in bubble properties from the prior static 
state ($R_0=9 m_0/4, V_0=0, T_0=8/(27\pi m_0),
\rho_{g0}, \rho_{s0}, \rho_{\tau0})$ (\ref{rho_g_id}-\ref{rho_t_id}),
to the ``initial'' conditions $(R_i,m_i,V_i,T_i,\rho_{gi}, \rho_{si}, \rho_{\tau i})$
for the subsequent relaxation to the final equilibrium state 
$(R_f,m_f,V_f=0,T_f,\rho_{gf}, \rho_{sf}, \rho_{\tau f})$ as $\tau\rightarrow\infty$.
We find 
\bea
R_i &=& R_0, \\
V_i &=& -\frac{9\pi Q_{L0}}{Q_{L0}-Q_{R0}} \bar{A}, \\
\bar{m}_i &=& \bar{m}_0 + \frac{27 \pi \bar{m}}{4} \bar{A}, \\
\bar{T}_i &=& \bar{T}_0 - \frac{9}{2 (Q_{L0}-Q_{R0}) \bar{\tau}_u} \bar{A}, \\
\rho_{gi} &=& \rho_{g0} + \left( \frac{\ell}{\bar{m}} 
                               - \frac{27 \alpha \rho_{g0}}{2 \bar{T}_0 \bar{\tau}_u (Q_{L0}-Q_{R0})}\right) \bar{A},\\
\rho_{\tau i} &=& \rho_{\tau 0} + \frac{27 \alpha \rho_{g0}}{2 \bar{T}_0 \bar{\tau}_u (Q_{L0}-Q_{R0})} \bar{A},\\
\rho_{s i} &=& \rho_{s 0},
\eea
where $\bar{A}\equiv A \bar{m}/\ell$ and $\bar{T}\equiv{T}/\ell$. Next,
we assume the solution for $\tau>0$ can be written
as a superposition of the three linear modes found in Sec.~\ref{vac_sd},
plus a relevant constant particular solution to fully (in addition to
the amplitudes of the modes) account for the initial conditions.
Assuming we choose parameters ($\alpha,\beta,\tau_u$) to give a stable
bubble, plus some dissipation $\zeta_0$ to give a static
state at $\tau=\infty$, we can then straight-forwardly read off the
final state by evaluating this solution at $\tau=\infty$. Of
particular relevance here is $R_f/m_f$, which in the large mass limit we find to be
\be
\frac{R_f}{m_f}  = \frac{3 (15\alpha-4)/4 - 4\sqrt{3}/(9\bar{m})}
                   {6\alpha-\beta-2} + O(\bar{m}^{-2}).
\ee
The linear mode analysis {\em assumed} what we want, namely that $R_f/m_f=9/4$, 
so for consistency here this becomes a constraint:
\be 
\alpha=2/3 + \beta -\frac{16\sqrt{3}}{81\bar{m}} + O(\bar{m}^{-2}). \label{alpha_fix}
\ee
Intriguingly, this can be expressed as
\be
\alpha=\beta + \rho_{\tau 0} \frac{8 \pi}{\ell \sqrt{3}} + O(\bar{m}^{-2}).
\ee

\section{x(r,t) map}\label{map_rx_A}

We define the map between the metric $r$ and code $x$ coordinate
as follows. First define a quadratic map between
$x$ and an intermediate coordinate $\hat{r}$ via:
\bea
x(\hat{r},t) &=& a(t) \hat{r} + b(t) \hat{r}^2, \ \ \ \hat{r}\leq R(t)\\
             &=& c(t) + d(t) \hat{r} + e(t) \hat{r}^2, \ \ \ \hat{r}\geq R(t).
\eea
The functions $a(t),b(t),c(t),d(t),e(t)$ are easily solved for by
imposing the list of conditions given in Sec.~\ref{rx_map}, and that 
$x(\hat{r},t=0) = \hat{r}$. We then stretch the exterior
part of the map to give $r(\hat{r})$:
\bea
r(\hat{r})&=&\hat{r}, \ \ \ \hat{r} \leq R(t)\\
          &=&\hat{r} + \hat{r}_{out} (R_s -1) \left[\frac{\hat{r}-R(t)}{\hat{r}_{out}-R(t)}\right]^3, \nonumber\\
          & & \ \ \ \hat{r}\geq R(t),
\eea
where the constant parameter $R_s$ controls how far away in 
$r$ we want the outer boundary location $\hat{r}_{out}=x_{out}$ to be.
For the back-reacting examples presented in Sec.~\ref{diss_EKG_results}
we used $R_s=10$, and $R_s=40$ for the non-back-reacting cases.

\section{Weak-form integration stencil}\label{sec_wfs}

We integrate the evolution equations (\ref{Ct_eom}-\ref{psi_eom})
about the location of the singular surface layer using the method outlined in Sec.~\ref{sec_weak}.
We use the map described in the previous section to keep it at a constant
coordinate location $x_0$, and if necessary adjust the initial position of the shell
to make sure $x_0$ coincides exactly with a vertex $i_0$ of the mesh.
We use a two cell wide piecewise linear test
function
\bea
v(x)&=&1 + \frac{(x-x_0)}{\Delta x}, \ \ \ x_0-\Delta x \leq x \leq x_0, \\
    &=&1 + \frac{(x_0-x)}{\Delta x}, \ \ \ x_0 \leq x \leq x_0 + \Delta x, \\
    &=&0, \ \ \, elsewhere,
\eea
where $\Delta x$ is the mesh spacing. Similarly, we decompose all metric and scalar field functions
in a basis of piecewise linear functions in $x$,
assuming the exact values are stored at grid vertices. For example,
\be
f(x)= f_{i-1}\frac{x_i-x}{\Delta x} + f_i \frac{x-x_{i-1}}{\Delta x}\ \ \ x_{i-1}\leq x \leq x_i,
\ee
where the notation $f_i$ means $f(x=x_i)$, with $x_i \equiv  i \Delta x$.
For a quantity $f$ that is discontinuous, hence multi-valued at $i_0$, we will
use the notation $f_L$ ($f_R$) to denote its value just to the left (right)
of $i_0$.
We can then analytically integrate (\ref{iqle}), arriving at an algebraic
equation that we can solve for the time derivative of the quantity of interest at $i_0$. 
Note that if one wanted to increase the accuracy of the scheme at the surface
layer one could do so by using higher degree polynomials (or other, smoother
basis functions) to represent the fields and test function.

Before writing down the resultant stencil, we note a couple of technical complications 
to reach the equivalent of the final integral given in (\ref{iqle}), related 
to our dual $r,x$ coordinate scheme.
The first is we are integrating in $x$, so need to include the Jacobian
of the coordinate transformation in the integral (\ref{svi}), and carry it through
the subsequent integration by parts. Second, our Runge-Kutta integration
scheme requires $\partial f(x,t)/\partial t$ at fixed $x$, though all
the time derivatives in (\ref{Ct_eom}-\ref{psi_eom}) are at fixed $r$; hence
we also need to transform between $\dot{f}$ and $\partial f(x,t)/\partial t$
(recall our notation $\dot{f} \equiv \partial f(r,t)/\partial t$,
$f' \equiv \partial f(r,t)/\partial r$). 

With all that, our first order accurate finite volume form of (\ref{qle}), which we repeat here
for reference:
\be
\dot{f}(t,r)-g'(t,r)+h(t,r)+\delta(r-R) S(t,r) = 0,
\ee
can be written as
\bea
\frac{d}{dt} f_{i_0}
      =  &-& \frac{1}{4}[g' + f'\cdot r_t]_{i_0+1} + \frac{1}{4}[g' + f'\cdot r_t]_{i_0-1} \nonumber \\
         &-& \frac{1}{4}[f\cdot r_{tr} + (f\cdot r_t +g) x''\cdot r_x]_{i_0+1} \nonumber\\
         &+&    \frac{1}{4}[f\cdot r_{tr} + (f\cdot r_t +g) x''\cdot r_x]_{i_0-1} \nonumber\\
         &-& \frac{1}{2}(h_L + h_R) - \frac{1}{2}(f_L+f_R)\cdot r_{tr} \nonumber \\
         &-& \frac{1}{2}[(f_L\cdot r_{t,i_0} +g_L)\cdot x''_L \nonumber\\
         & & \ \ \    +(f_R\cdot r_{t,i_0} +g_R)\cdot x''_R]\cdot r_{x,i0} \nonumber\\
         &+& \frac{3}{4\Delta x}\bigg[ (g_R-g_L+(f_R-f_L)\cdot r_t)\cdot x'_{i_0} \nonumber\\
         &\ &\ \ \ \ \ \ \   + [(g+f\cdot r_t)\cdot x']_{i_0+1}\nonumber\\
         &\ &\ \ \ \ \ \ \   - [(g+f\cdot r_t)\cdot x']_{i_0-1}\nonumber\\
         &\ &\ \ \ \ \ \ \   - 2 S_{i0}\cdot x'_{i0} \bigg],
\eea
where
$r_t\equiv \partial r(x,t)/\partial t$,  $r_x\equiv \partial r(x,t)/\partial x$,
$r_{tr} \equiv [\partial^2 r(x,t)/(\partial t \partial x)]\cdot \partial x(r,t)/\partial r$.
This elevates to a second order accurate scheme when $S_{i0}\rightarrow 0$, and
all ``L'' values equal their ``R'' value neighbours.

\section{Rate of Energy Loss from AdS interior}\label{app_loss_rate}
In light-like coordinates, the line element for AdS is:
\begin{equation}
ds^2 = \cos^{-2}(\ell r/\sqrt{3} ) (-dt^2 + dr^2) + 3/\ell^2 \tan^2(\ell r/\sqrt{3}) d\Omega^2
\end{equation}
which we distinguish from the ``standard coordinates'' ($r,R$) which give
\begin{equation}
ds^2 = -(1+R^2\ell^2/3) dt^2 + (1+R^2\ell^2/3)^{-1} dr^2 + R^2 d\Omega^2
\end{equation}

General solutions to scalar field propagation in AdS in light-like coordinates can be expressed
as a superposition of modes given by
\bea
\Phi_j(t,x) &=&  d_j \cos( \omega_j \ell t/\sqrt{3})  \cos^3(\ell r/\sqrt{3})\nonumber\\
            & & \, {}_2F_1(-j,3+j,3/2,\sin^2(\ell r/\sqrt{3}))\, ;
\eea
with $\omega_j^2 = (3+2j)^2 \ell^2/3$ and $d_j=4 \sqrt{(j+1)(j+2)/\pi}$ ($j\in 0,1,..$). Such modes are orthonormal
and complete as $\ell \rightarrow \infty$. For our regime of interest then we can use this (quasi) basis, for
large $\ell$, to describe the exterior pulse once it enters the AdS region. In particular, we are interested
in the reduction of (the interior) energy ($E$) within the AdS. To this end, we can make use of the analysis presented in~\cite{Calabrese_2003} 
to reach the intuitive result of
\begin{equation}
E_{,t} = 4 \pi (V_+^2 - V_-^2) R_{o}^2
\end{equation}
with $V_{+,-} = \pm \alpha u^a \partial_a \Phi + \alpha D$ the incoming (outgoing) modes of the solution at the outer boundary $r=R_o$;
$u^a$ is the unit timelike normal at $R_o$, $D = \gamma^{ij} n_i \partial_j \Phi$, and $\alpha=1/\cos(\ell r/\sqrt{3})$.
At such a boundary, the AdS region looses energy through $V_-$ but does not gain energy through $V_+$ as little ``comes back''
from the exterior region. We can thus take it to zero, so energy is lost at a rate $E_{,t} = - 4 \pi V_-^2 R_{o}^2$. 
We can then replace $\Phi$ in terms of its normal
modes; it is clear higher modes will reduce the energy more  effectively than the lowest one $\omega_0$. Said
differently, energy supported by higher frequency modes leaks out at a faster rate out of the AdS region. The long-term behavior
is given by the lowest mode, and the energy loss within 1 period of oscillation is, to leading order in $1/\ell$,
\begin{equation}
\Delta E \propto - A_0^2 \pi^2 \frac{1}{m^2 \ell^3}
\end{equation} 
(with $A_0$ the amplitude of the mode).
We can then use a ``quasi-adiabatic'' argument to say an amount of energy 
\begin{equation}
\Delta E \propto - A_0^2 \pi^2 \frac{1}{m^2 \ell^2} \Delta T
\end{equation}
is lost in the interior region over the period $\Delta T = (2 \pi)/(\ell \sqrt{3})$. Now, the energy within
the AdS region is $\propto A_0^2/\ell$, so $dE/dt \propto 2 A_0/\ell dA_0/dt$ and we can
use $\Delta E/\Delta T$ to approximate the left hand side to arrive at,
\begin{equation}
\frac{dA_0}{dt} \propto - \frac{\pi A_0}{m^2 \ell^2} \ell
\end{equation}
and so, $A_0(t) \approx \exp(-p t \ell)$ with
\begin{equation}
p \approx - \frac{\pi}{m^2 \ell^2}.
\end{equation}
Thus the energy decays as $E(t) \propto \exp(-2 p t \ell)/\ell$, and so $\log E \approx - \frac{2 \pi}{ m^2 \ell} \, t$.
Notice the above expression is with respect to time measured at the origin of AdS, which is related
to the asymptotic time $t_a$ by $t \approx t_a/(m \ell)$. Consequently, $\log E \approx - \frac{2 \pi}{ m^3 \ell^2} \, t_a$. This
behavior is consistent with the results shown in figures ~\ref{fig:SFE} and ~\ref{fig:qnm12}.
For reference, we can now explore the associated timescale for this energy to leak out of AdS and become an ``observable signature'' in the 
asymptotically flat (AF) region.
The timescale is given by $\tau_D \simeq m^3 \ell^2$; taking $\ell = 1/L$ with $L$ a lengthscale and assuming $m=10^q M_{\odot}$, one has
\begin{equation}
\tau_D = 10^{3q} (\mbox{m}/L)^2 \mbox{s}
\end{equation}
For instance, for $L=L_{\rm{Planck}}\simeq 10^{-35}\mbox{m}$ and $q=1$, $\tau_D = 10^{53} \mbox{s} \approx 10^{56} t_{\rm{Hubble}}$. Requiring instead
that $\tau_D \simeq t_{\rm{Hubble}}$ or $\tau_D \simeq 1\mbox{yr}$, $L$ should be $\simeq 10^{-7}, 10^{-4}$m respectively.

As a last remark, we can employ a similar argument to explore what takes place at early times. When a pulse with a given
frequency $\omega_{\rm{AF}}$ in the AF region, begins to fall in the AdS region, its frequency would be blueshifted to $\omega_i \simeq \omega_{\rm{AF}} m \ell$,
and would be supported, in terms of the AdS modes, by  a spectra of (almost) normal frequencies given by $\omega_j = \pm (3+2j) \ell/\sqrt{3}$. Thus,
the pulse would be described by the same modes in a way that is largely insensitive to the scale determined by $\ell$. For a mode with index $j$
the above timescale estimate results in
\begin{equation}\label{mode_jL}
\tau_D = 10^{3q} j^{-2} (\mbox{m}/L)^2 \mbox{s}
\end{equation}
indicating the AdS could help potentially render microscopic $j$ scales into significantly longer ones for higher values of $j$. Of course, 
this depends on the content of the pulse in the AF region. Rough estimates however imply not very high $j$'s are encountered with significant strength
for the relatively simple frequency content of waves driven by a quasi-circular merger.

\bibliography{references}
\bibliographystyle{ieeetr}
\end{document}